\documentclass[final,times,twocolumn]{elsarticle}

\biboptions{sort&compress}

\usepackage{amssymb}
\usepackage{amsmath}

\usepackage{xcolor} 
\usepackage{soul}
\usepackage{stfloats}

\usepackage{natbib}
\usepackage{siunitx}
\usepackage{hyperref}

\DeclareSIUnit{\litre}{l}

\usepackage[mathlines,modulo,switch]{lineno}

\usepackage{caption}
\journal{Journal of Non-Newtonian Fluid Mechanics}

\graphicspath{{figures/}}

\newcommand{\eqnref}[1]{Eq.~(\ref{#1})}
\newcommand{\figref}[1]{Fig.~\ref{#1}}

\begin{document}

\begin{frontmatter}



\title{Tuning fluid frictional drag through flow-induced aggregation \\ of wormlike micelles} 


\author[1]{Lucas Warwaruk\corref{cor1}} 

\author[2]{Amanda Tiano}

\author[2]{Berith Isaac}

\author[1]{Gareth H. McKinley\corref{cor2}}

\affiliation[1]{organization={Hatsopoulos Microfluids Laboratory, Department of Mechanical Engineering, Massachusetts Institute of Technology},
            addressline={33 Massachusetts Ave}, 
            city={Cambridge},
            postcode={02139}, 
            state={MA},
            country={United States}}

\affiliation[2]{organization={Department of Biology, Brandeis University},
            addressline={415 South St}, 
            city={Waltham},
            postcode={02453}, 
            state={MA},
            country={United States}}

\cortext[cor1]{E-mail address: warwar95@mit.edu (L. Warwaruk).}
\cortext[cor2]{E-mail address: gareth@mit.edu (G. H. McKinley).}

\begin{abstract}

Charged amphiphilic surfactants are versatile polymorphic molecules which can form supramolecular structures, such as wormlike micelles. We study the rheology and non-Newtonian fluid dynamics of the cationic surfactant trimethyltetradecylammonium chloride (TTAC) combined with the aromatic counter-ion sodium salicylate (NaSal). In the non-dilute regime, this solution exhibits a linear viscoelastic response consistent with a single-mode Jeffreys model, and in capillary pinch-off experiments, fluid filaments form extended cylindrical threads, indicating strong strain-stiffening. In the dilute regime the linear viscoelasticity and extensional rheology become immeasurably small, but the TTAC/NaSal mixture develops flow-induced structures at high shear rates, as the result of an end-to-end concatenation of micelles produced by shear-induced flow alignment. These flow-induced structures augment both the measured viscosity and the viscous drag in laminar flows at high shear rates (by as much as 50~\%). In turbulent Taylor-Couette flow, the flow-induced structures produce a sudden and dramatic destabilizing influence on the stable toroidal Taylor vortices, resulting in upwards of a 200~\% enhancement in turbulent drag. However, with increasing Reynolds number, the turbulent flow transitions smoothly to a state of drag reduction, generating up to a 30~\% reduction in fluid frictional drag but without the shear-induced degradation that plagues drag reduction strategies based on solutions of high molecular weight polymer, such as polyacrylamide. These results show that surfactants are a unique and effective modifier of fluid frictional drag in both laminar and turbulent flow regimes, capable of producing both significant levels of drag enhancement and drag reduction by tuning the composition and operating conditions.

\end{abstract}




\end{frontmatter}



\section{Introduction}

Adding a small amount of high-molecular weight polymer to a turbulent liquid wall-bounded flow can produce an appreciable reduction in fluid frictional drag (upwards of 60~\%). Since the first reports of polymer drag reduction by \citet{Toms1948} nearly 80 years ago in the first International Congress on Rheology, the mechanics of polymer drag reduction has been extensively studied; for example, see the many review articles relating to polymer drag reduction by \citet{Lumley1969,Hoyt1972,Virk1975,White2008,Graham2014,Xi2019} and others. Over its 50 year history, the \textit{Journal of Non-Newtonian Fluid Mechanics} (JNNFM) has also published many seminal contributions relating to the topic of polymer drag reduction, including detailed works that underpin the mechanism for drag reduction \citep{Renardy1995,Yarin1997,Larson2002}, foundational investigations on how to simulate high-Reynolds number viscoelastic flows \citep{Orlandi1995,Dimitropoulos1998,Li2006,Iaccarino2010}, high-fidelity experimental characterizations of polymer-laden turbulence \citep{oliver1983,Pinho1990,Vlassopoulos1993,Escudier1999,Escudier2009,Pereira2013}, and many more. 

Dilute solutions of high molecular weight polymers have been successfully used in a variety of industrial applications for reducing fluid frictional drag. In the Trans-Alaska-Pipeline System, long-chain polymers are used to improve crude oil delivery by approximately 300~000 barrels per day; corresponding to approximately a 30~\% enhancement in pipeline throughput \citep{Burger1980,Burger1982}. Although dilute solutions of polymer additives greatly reduce turbulent drag their long-term performance is hindered by mechanical degradation. When subjected to strong fluctuating flows and large material strains, scission of the polymer chains causes the level of drag reduction to deteriorate over time \citep{Brostow1983}. Therefore, polymers are used almost exclusively in \textit{once-through} flow applications, such as fire suppression or municipal sewage systems, that mitigate the exposure of polymers to the high strain rates of pump impellers or flow constrictions \citep{Sellin1980,Figueredo2003}. However, for \textit{closed-loop} flow applications, such as recirculating liquid heating and cooling systems, drag-reducing additives that do not degrade with shear history are required.

Self assembling surfactant systems are another well-known class of drag-reducing additive, and they do not permanently degrade when subjected to high strain rates or large levels of accumulated strain \citep{Zakin2017}. The first documented report of surfactant drag reduction appeared in a patent by \citet{Mysels1949}, only one year after the discovery of polymer drag reduction by \citet{Toms1948}. 
Although some investigations have shown that self assembling surfactant systems can produce higher amounts of turbulent drag reduction than polymers \citep{Chara1993,Zakin1996}, surfactant-based fluids have generally garnered less wide-spread industry adoption and scientific attention. That said, there have been many key investigations that have characterized the microstructure, rheology, and drag reducing capacity of surfactant-based systems, many of which have also been published in \textit{JNNFM} over the past 50 years \citep{Lu1997,Lin2000,Gasljevic2001,Qi2003,Ge2008,Tamano2015}.

Surfactants are amphiphilic molecules, consisting of a hydrophilic head group and a hydrophobic tail \citep{Israelachvili2011}. The most effective drag-reducing surfactants investigated to date are cationic quaternary ammonium salts \citep{Qi2002}. When mixed with an anionic counter-ion (e.g., sodium salicylate), the surfactant molecules pack together to form supramolecular associations known as \textit{micelles} \citep{Israelachvili2011}. The equilibrium conformation of these micelles depend on a number of physicochemical conditions, including temperature, ionic strength, and surfactant concentration \citep{Larson1999}. Wormlike micelles that grow linearly and ultimately attain a similar contour length as drag-reducing polymers of $\mathcal{O}(1~$\SI{}{\micro\meter}), are regarded as being the ideal micellar microstructure for turbulent drag reduction \citep{Qi2002}. However, unlike polymers, wormlike micelles are \textit{self-reparable} and undergo reversible scission in strong deformations. Hence, wormlike micellar fluids are often colloquially referred to as \textit{living polymers} \citep{Larson1999}. 

The design requirements for drag-reducing polymer solutions are to achieve appreciable viscoelasticity with a low shear viscosity, and a relatively long relaxation time \citep{Owolabi2017,Warwaruk2025}. Therefore, most drag-reducing polymer solutions are often formulated to operate in a dilute concentration regime (with negligible polymer-polymer interactions) and consist of linear flexible macromolecules with an ultra-high molecular weight and a long Zimm relaxation time of $\tau_Z > 1~\mathrm{ms}$, that exceeds the Kolmogorov time-scale of viscous dissipation in the turbulent flow \citep{Pope2000,Xi2019}. Two synthetic polymers commonly used for drag reduction are polyethylene oxide and polyacrylamide, both of which can be synthesized with molecular weights exceeding $M_w~>~1~\mathrm{MDa}$ \citep{Xi2019}. However other naturally-sourced biopolymers derived from okra, aloe vera, flax seeds and even human saliva have been shown to reduce turbulent drag \citep{Coelho2016,Soares2019,Rajappan2019,Warwaruk2026b}. When a few parts-per-million (ppm) of any of these polymers are dissolved in water they disperse readily and the chains adopt a random coil configuration under rest conditions. The highly extensible nature of the flexible polymer backbones result in long relaxation times of $\tau_E > 1~\mathrm{ms}$ when the chains are elongated using capillarity-driven extensional rheology \citep{Owolabi2017}. The low viscosity and high extensibility of such fluids result in reductions of skin-friction drag by up to 70~\% in turbulent pipe, channel and boundary layer flows \citep{Warholic1999,Ptasinski2001,White2004}. Direct numerical simulations (DNS) of high Reynolds number wall-bounded turbulent flows utilizing Hookean or FENE (finitely extensible non-linear elastic) dumbbell constitutive models to describe the dynamical evolution in polymer-driven elastic stresses also produce similar levels of drag reduction as those observed in experiments \citep{Ptasinski2001,Min2003,Min2003a}. In these simulations the polymer solution is modeled as a dilute non-interacting suspension of highly extensible dumbbells by selecting a solvent-to-zero shear viscosity ratio ($\beta = \eta_s/\eta_0$) that is close to unity (e.g., $0.9 \lesssim \beta <1$), and then imposing a flow characterized by a large Weissenberg number $Wi = \tau_Z \dot{\gamma} \gg 1$ \citep{Min2003a}. Here the Weissenberg number is a non-dimensional measure of flow strength that represents the degree to which polymers are extended within the flow, and $\dot{\gamma}$ is a representative strain rate of interest in the flow \citep{Owolabi2017}.

Similar to polymers, drag-reducing solutions of surfactants have been shown to exhibit viscoelasticity and an enhanced extensional resistance (as measured using an opposed jet extensional rheometer or flow through a hyperbolic contraction)  \citep{Prudhomme1994,Wunderlich1987}. However, one unique property of dilute surfactant solutions is their ability to form \textit{shear-induced structures} \citep{Ohlendorf1986,Bewersdorff1988}. Above a critical shear rate, short semiflexible or rodlike micelles align and concatenate end-to-end to form longer supramolecular assemblies. In measurements of steady shear rheology, shear-induced structures cause the measured viscosity to be enhanced, resulting in an apparent shear-thickening. When measured in a stress-controlled up/down flow sweep this increase in viscosity is often found to be hysteretic, producing a re-entrant flow curve and exhibiting a multiplicity of accessible shear rates for a given stress. This is often associated with the formation of bands of different dynamical states along the vorticity direction of the imposed flow field (i.e., \textit{vorticity banding}) \citep{Dutta2018}. Although many early investigations have observed the formation of supramolecular structures in shear flow, their formation is not exclusive to shear deformation and can also be initiated using other sufficiently strong flow kinematics (e.g., extensional flows at $Wi>1$) that induce flow alignment and concatenation of microstructural units \citep{Wunderlich1987,Vasudevan2010,Recktenwald2019}. Hence, the supramolecular micellar structures formed during strong flows at $Wi>1$ are more correctly referred to as \textit{flow-induced structures} that are agnostic to the specific kinematics of the flow, and we use this nomenclature henceforth. 

To model the flow-induced structures of rodlike micelle solutions, Graham and coworkers \citep{Dutta2018,Hommel2021} proposed a new constitutive equation, the \textit{reactive rod model}, which is capable of producing the apparent shear and extensional flow thickening commonly seen in steady shear and extension. In inertia-free cylindrical Couette flow, \citet{Hommel2024} also demonstrated that a heterogeneous shear rate (induced by curvature) can contribute to spatial gradients in the length of the resulting supramolecular microstructure that generate finger-like flow instabilities. Interestingly, polymers have also been shown to induce an apparent shear-thickening in cylindrical Couette flow at high shear rates, through the initiation of elastically-driven curved streamline instabilities \citep{Steinberg1998,Larson1992}. In low Reynolds number flows with negligible levels of inertia, these elastic instabilities give rise to a new spatio-temporally fluctuating flow state christened \textit{elastic turbulence}, or ET \citep{Groisman2000,Steinberg2021}. Thus, although surfactants and polymers can both drive flow instabilities at vanishingly small Reynolds numbers, the viscoelastic mechanism for triggering flow instabilities using dissolved polymers is different than that associated with FIS formation in self-assembling surfactants.

At high Reynolds numbers, when the fluid flow is dominated by inertia, dilute polymer solutions can also drive elasto-inertial instabilities that lead to another chaotic flow state coined \textit{elasto-inertial turbulence} or EIT \citep{Samanta2013,Castillo2022,Dubief2023}. The connections between ET, EIT and drag reduction at high Reynolds number are still being explored \cite{Xi2019,Dubief2023}. However, it has been argued that EIT is connected to the asymptotic state of maximum drag reduction (MDR) \citep{Choueiri2018}. At high Reynolds numbers ($Re\gg1000$), the level of drag reduction achieved in wall-driven turbulence does not exceed a limiting asymptote (MDR) despite further supplementing the polymer concentration or increasing the molecular weight \citep{Virk1970}. Flows at MDR are not laminar, but exhibit weakened spatiotemporal fluctuations and significant elastic stresses (similar to EIT) \citep{Warholic1999}. 

Although the transition to turbulence in dilute solutions of flexible polymers has been heavily investigated, surfactant-laden flows have garnered significantly less attention. \citet{Hommel2024} demonstrated that flow-induced structures comprised of rodlike micelles can trigger flow instabilities in cylindrical Couette flow at vanishingly small Reynolds numbers. However, a connection between the formation of flow-induced structures and turbulent drag reduction at large Reynolds numbers has yet to be made. The recent experimental investigation of \citet{Horimoto2025} demonstrated that the surfactant cetyltrimethylammonium chloride (CTAC) can be used to modify drag in a co- and counter-rotating Taylor-Couette flow. In a differential co-rotating Taylor-Couette flow (with a ratio of $R_i/R_o=0.732$ between the inner and outer walls) \citet{Horimoto2025} showed that a 100~ppm solution of CTAC produced an incredibly large \textit{enhancement} in turbulent drag of 700~\% at a Reynolds number of $Re_i = 50~000$ (based on the angular velocity of the inner cylinder). However, in counter-rotation, the same CTAC solution produced close to a 50~\% \textit{reduction} in turbulent drag at the same Reynolds number. They attribute the drag enhancement observed in the CTAC solutions to EIT (arising from viscoelastic rheology), similar to the mechanism for drag enhancement (inertio-elastic G\"{o}rtler vortices) observed in DNS of Taylor-Couette flow utilizing viscoelastic FENE dumbbell constitutive models \citep{Liu2013,Song2019}. However, \citet{Horimoto2025} acknowledged that direct measurements of the viscoelastic rheology were challenging and could not be reliably performed due to the dilute nature of their CTAC solutions. They also did not present direct measurements of the shear rheology of their CTAC solutions, and it is unclear what role flow-induced structure formation might have in modifying drag in Taylor-Couette flow. Therefore, to our knowledge, no previous investigation has been able to identify the cause of drag modification in turbulent Taylor-Couette flow of dilute self-assembling surfactant solutions. 

In the present work we investigate the role of flow-induced structures in modifying the Taylor-Couette dynamics of dilute self-assembling surfactant solutions over a wide range of Reynolds numbers $370\leq Re \leq 99~000$, where the influence of inertia is not negligible. Although non-dilute solutions of the surfactant and counter-ion exhibit a high degree of viscoelasticity (attributed to continued creation and destruction of entanglements between the wormlike micellar structures), diluted solutions of the same composition have immeasurably small levels of elasticity (as measured with small amplitude oscillatory shear) and transient extensional rheology (using the Dripping-onto-Substrate method). The only observable non-Newtonian feature of these low viscosity dilute surfactant solutions is FIS formation at high shear rates as marked by an apparent shear thickening in measurements of steady shear rheology. In Taylor-Couette flow, we show that the flow-induced structures first produce an appreciable advective drag enhancement in the presence of inertial wavy Taylor vortices. At sufficiently large Reynolds numbers, nonlinear viscoelastic effects generated by these supramolecular structures drive the flow into a state of turbulent drag reduction -- providing, for the first time, a direct link between the formation of flow-induced structures and their ability to enhance and reduce turbulent drag with increasing Reynolds number. 

\section{Test Fluids}\label{sec:test_fluids}

There are many different types of surface active molecules \citep{Israelachvili2011,Larson1999}. The unique microstructures of surfactants enable an immense design space for building fluids with unique rheological properties. In the current section, we provide the reader with a high-level review of surfactant chemistry and how it motivated the formulation of the test fluids used in the present investigation. 

\begin{figure*}[b!]
\centering
\includegraphics[width=\textwidth]{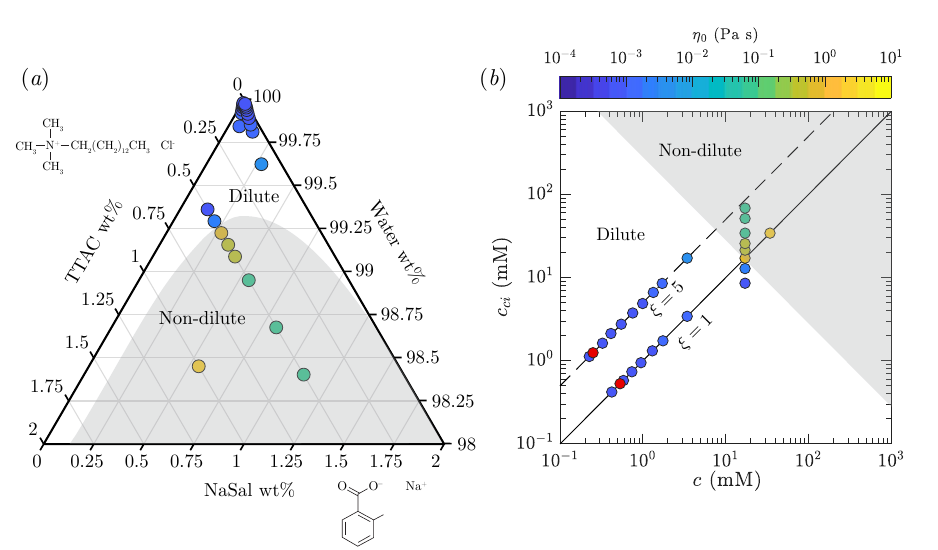}
\caption{A ternary diagram that shows the relative proportions of the surfactant trimethyltetradecylammonium chloride (TTAC), sodium salicylate (NaSal) and the solvent (deionized water) is demonstrated in (\textit{a}). The chemical structures of TTAC and NaSal are illustrated in (\textit{a}). A state diagram of the molar concentration of counter-ion, NaSal, versus the molar concentration of surfactant, TTAC, is shown in (\textit{b}). Here the molar ratio of NaSal to TTAC is denoted as $\xi = c_{ci}/c$. The solid and dashed black lines in (\textit{b}) represent $\xi =1$ and $\xi=5$, respectively. Markers are coloured according to the magnitude of the zero-shear--rate viscosity and mark the concentrations of TTAC and NaSal considered in the current investigation. The grey shaded regions in (\textit{a}) and (\textit{b}) represent the approximate regime in which the micelles begin to interact with one another and the solution viscosity increases markedly. Red markers denote the critical micelle concentration $c_{\mathrm{cmc}}$ for the solutions with molar ratios of $\xi=1$ and $\xi =5$. For a solution with no salicylate, i.e., $c_{ci} = 0~\mathrm{mM}$, the critical micelle concentration is $c_{\mathrm{cmc}}=5.5~\mathrm{mM}$ \citep{Durand2006}.}
\label{conc}
\end{figure*}

Although our interests in surfactants pertain to their ability to reduce turbulent drag, surfactants are more commonly used to modify surface tension $\varGamma$ along an immiscible liquid-vapour or liquid-liquid interface; the word surfactant is a \textit{portmanteau} (blend) of the phrase \textit{surface-active-agent}. When dissolved in water with an interface, surfactants congregate along the interface due to their inherent amphiphilic chemical structure, and reduce the apparent surface tension \citep{Larson1999}. If the surfactant concentration along the interface is inhomogeneous, the resulting gradients in surface tension can drive Marangoni flows along the interface \citep{Manikantan2020}; however, this is beyond the scope of the present investigation.

As the concentration of surfactant increases, the surface tension $\varGamma$ of the solution will typically decrease. Above a critical concentration, surfactants fully saturate the interface and the surface tension will no longer change with increasing surfactant concentration \citep{deGennes2004}. At these high concentrations, the remaining surfactants aggregate in the bulk liquid phase to produce growing micelles; therefore, the critical concentration is also referred to as the \textit{critical micelle concentration}, herein denoted as $c_{\mathrm{cmc}}$. The critical micelle concentration, as well as the shape of the resulting micellar microstructure depend on a number of factors including the type of surfactant, temperature and salinity \citep{Larson1999}.

Surfactant molecules can be broadly classified based on the charge of their hydrophilic head group: anionic (negatively charged), cationic (positively charged), zwitterionic (with zero net charge, but both negative and positively charged components) or nonionic (uncharged) \citep{Qi2002}. Any one or combination of these surfactants can be used to generate drag-reducing fluids; however, the most commonly used drag-reducing surfactant formulation is cationic, for a number of practical reasons discussed in \citet{Qi2002}. A common type of cationic surfactant is an alkyltrimethylammonium surfactant with a generic chemical structure of $\mathrm{C}_n \mathrm{H}_{2n+1} \mathrm{N}^+(\mathrm{C}\mathrm{H}_3)_3 \mathrm{X}^-$, where $n$ is the number of carbon atoms in the hydrophobic tail (or alkyl chain) of the surfactant molecule and $\mathrm{X}^-$ is a negatively charged counter-ion (e.g., chloride, $\mathrm{Cl^-}$; bromide, $\mathrm{Br^-}$, etc.). 

Wormlike micelles consisting of cationic surfactants are more easily formed using aromatic negatively-charged counter-ions, such as salicylate, $\mathrm{Sal}^-$ \citep{Qi2002}. When dissolved with the surfactant, these aromatic molecules are able to intercalate between the heads of neighboring surfactant molecules tightly, resulting in the formation and growth of long wormlike micelles. As the concentration of the aromatic counter-ion $c_{ci}$ increases, the critical micelle concentration decreases and the length of the wormlike micelles grows. At sufficiently high concentrations, the resulting solutions are generally viscoelastic, shear thinning, and also show extensional strain stiffening \citep{Prudhomme1994,Wunderlich1987}.

In the present investigation we consider the cationic surfactant trimethyltetradecylammonium chloride, herein referred to as TTAC, with $n=14$ carbon atoms in its hydrophobic tail, a chloride counter-ion $\mathrm{X}^{-} = \mathrm{Cl}^-$, and a molar mass of $M = 291.94~\mathrm{g~mol}^{-1}$. Solid TTAC powder (87212, MilliporeSigma) is dissolved in deionized water at a concentration of $c = 69$~mM (or mmol~l$^{-1}$). \citet{Durand2006} have reported that the critical micelle concentration for TTAC dissolved in water (at $T = 298~\mathrm{K}$) is $c_{\mathrm{cmc}} = 5.5~\mathrm{mM}$. Without a strongly binding counter-ion, e.g., salicylate, TTAC forms spherical micelles at concentrations $c\geq c_{\mathrm{cmc}}$ \citep{Durand2006}. To promote the formation of wormlike micelles, TTAC is mixed with sodium salicylate NaSal (molar mass $M = 160.10~\mathrm{g~mol}^{-1}$). A separate solution was prepared by adding solid sodium salicylate powder (S3007, MilliporeSigma) to deionized water at a concentration of $c_{ci} = 156$~mM. The two concentrated solutions of TTAC and NaSal are then mixed at different proportions, where the molar ratio of counter-ion to surfactant in solution is denoted as $\xi = c_{ci}/c$. All test fluids are maintained at the same temperature of $T=298~\mathrm{K}$. Changes in temperature can also have a significant impact on the microstructure of the resulting micelles; however, the influences of temperature are beyond the scope of the present investigation.

In \figref{conc} we provide an overview of the different concentrations of TTAC and sodium salicylate considered in the present investigation, represented in the form of a ternary composition diagram with the relative amounts of TTAC, NaSal, and solvent contained in the test solutions studied. Markers are coloured according to their measured zero-shear--rate viscosity $\eta_0$. Shear viscosity data was collected using a torsional rheometer (MCR 703, Anton Paar) equipped with a parallel plate geometry (using a plate radius of either $R = 12.5~\mathrm{mm}$ or $21.5~\mathrm{mm}$) for more viscous fluids and a double gap concentric cylinder geometry (DHR-3, TA Instruments), with an average gap of $d = 0.45~\mathrm{mm}$, for low viscosity samples. These viscosity measurements will be shown and discussed in greater detail below. For now, \figref{conc}(\textit{a}) shows that solutions with TTAC in excess of $0.5~\mathrm{wt\%}$, but also with high concentrations of sodium salicylate (above $0.25~\mathrm{wt\%}$), exhibit a significantly larger viscosity compared to the solvent (i.e., $\eta_0(c;\xi) \gg \eta_s$, where the viscosity of water is $\eta_s = 0.89\times10^{-3}~\mathrm{Pa~s}$). An approximate region that delineates the viscous solutions with $\eta_0\geq10\eta_s$, in a non-dilute regime, from those with a low viscosity similar to the solvent ($\eta_0<10\eta_s$), is shown in grey in \figref{conc}(\textit{a}). 

\figref{conc}(\textit{b}) presents a similar phase diagram as \figref{conc}(\textit{a}), but instead showing the molar concentration of sodium salicylate $c_{ci}$ plotted with respect to the surfactant concentration $c$ for the different test solutions. For the dilute drag-reducing surfactant solutions, measurements are performed using a range of TTAC concentrations, but for two constant molar ratios of NaSal to TTAC, $\xi=1$ and $\xi=5$, shown by the black solid and dashed lines in \figref{conc}(\textit{b}), respectively. Although all of the dilute solutions exhibit a low viscosity that is similar in magnitude to water, nearly all solutions are above the critical micelle concentration, denoted using the red symbols in \figref{conc}(\textit{b}) for the molar ratios of $\xi = 1$ and $\xi = 5$.

The critical micelle concentration $c_{\mathrm{cmc}}$ is determined for the TTAC solutions with different counter-ion molar ratio, $\xi = 1$ and $\xi = 5$, by measuring the surface tension $\varGamma$ of the solutions with decreasing concentration $c$. Surface tension of the liquids are measured using a force tensiometer (DCAT11, DataPhysics) equipped with a platinum Du No\"{u}y ring (RG11, DataPhysics), and using a ring push/pull method.

\begin{figure}[b!]
\centering
\includegraphics{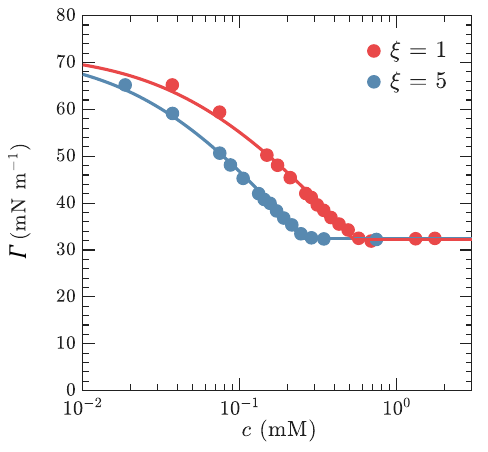}
\caption{Liquid-air surface tension $\varGamma$ as a function of TTAC concentration $c$ for two different counter-ion molar ratios, $\xi$. Solid lines are fits to $\varGamma(c)$ using the Szyszkowski equation of \eqnref{eqn:SFT}. Twenty measurements (ten ring pushes, and ten pulls) are collected for each data point. Measurement uncertainty is comparable to the marker size.}
\label{CMC}
\end{figure}

\figref{CMC} presents measurements of the surface tension as a function of TTAC concentration for the two different counter-ion molar ratios of $\xi = 1$ and $\xi = 5$. Measurements are fit to the Szyszkowski equation \citep{Rosen2012}

\begin{equation}
\varGamma(c) = \begin{cases} 
      \varGamma_0-\frac{k_BT}{a_0}\ln(1+Kc) & \text{if } c < c_{\mathrm{cmc}} \\
      \varGamma_{\infty} & \text{if } c \geq c_{\mathrm{cmc}}
   \end{cases}
   \label{eqn:SFT}
\end{equation}

\noindent
where $\varGamma_0 = 72.0~\mathrm{mN~m^{-1}}$ is the surface tension of the solvent (water), $\varGamma_{\infty} = 32.5~\mathrm{mN~m^{-1}}$ is the surface tension for high concentrations $c\geq c_{\mathrm{cmc}}$, $a_0$ (units of $\mathrm{m}^2$) is the effective area occupied by surfactant molecules at the liquid-air interface, and $K$ (units of $\mathrm{mM^{-1}}$) is an adsorption coefficient \citep{Zdziennicka2012}.

The measurements of surface tension as a function of concentration $\varGamma(c)$ presented in \figref{CMC} show good agreement with fits to \eqnref{eqn:SFT}, represented by the solid coloured lines. For the TTAC solutions with a molar ratio of $\xi = 1$, the critical micelle concentration is $c_{\mathrm{cmc}} = 0.53~\mathrm{mM}$, the head packing area along the interface is $a_0 = 0.24~\mathrm{nm}^2$ and the adsorption constant is $K = 16.6~\mathrm{mM^{-1}}$. For the high molar ratio of $\xi = 5$, the critical micelle concentration is lowered, $c_{\mathrm{cmc}} = 0.25~\mathrm{mM}$, the head packing area along the interface is $a_0 = 0.22~\mathrm{nm}^2$ and the adsorption constant is $K = 28.1~\mathrm{mM^{-1}}$. We note that \citet{Durand2006} reported that TTAC without any added salicylate counter-ion had a critical micelle concentration of $c_{\mathrm{cmc}} = 5.5~\mathrm{mM}$. Increasing the molar concentration of salicylate ($\xi$) systematically reduces the critical micelle concentration $c_{\mathrm{cmc}}$.

\section{Non-dilute Solutions of Wormlike Micelles}\label{sec:semi-dilute}

Due to their low concentrations, rheological measurements of the dilute drag-reducing surfactant solutions are challenging. Therefore a thorough rheological characterization is first performed for the non-dilute micellar solutions, previously shown in \figref{conc}, at higher TTAC concentrations of $c=17~\mathrm{mM}$ and 34~mM. Steady shear rheograms and DoS (dripping-onto-substrate) capillarity-driven extensional rheology measurements are first shown for surfactant solutions with a fixed TTAC concentration of $c = 17~\mathrm{mM}$ ($5~\mathrm{g~l^{-1}}$), but different amounts of sodium salicylate. We then present images of the wormlike micelle microstructure, collected using cryogenic transmission electron microscopy, for a higher concentration system of $c = 34~\mathrm{mM}$ ($10~\mathrm{g~l^{-1}}$) with a molar ratio of $\xi = 1$. Lastly, measurements of the linear viscoelasticity and flow-induced birefringence are also presented for the same solution. This provides a comprehensive characterization of the non-dilute solution rheology, before we then consider the dilute solutions relevant for turbulent drag reduction. 

\subsection{Influence of counter-ion Concentration}\label{sec:ion}

\begin{figure*}
\centering
\includegraphics[width=0.95\textwidth]{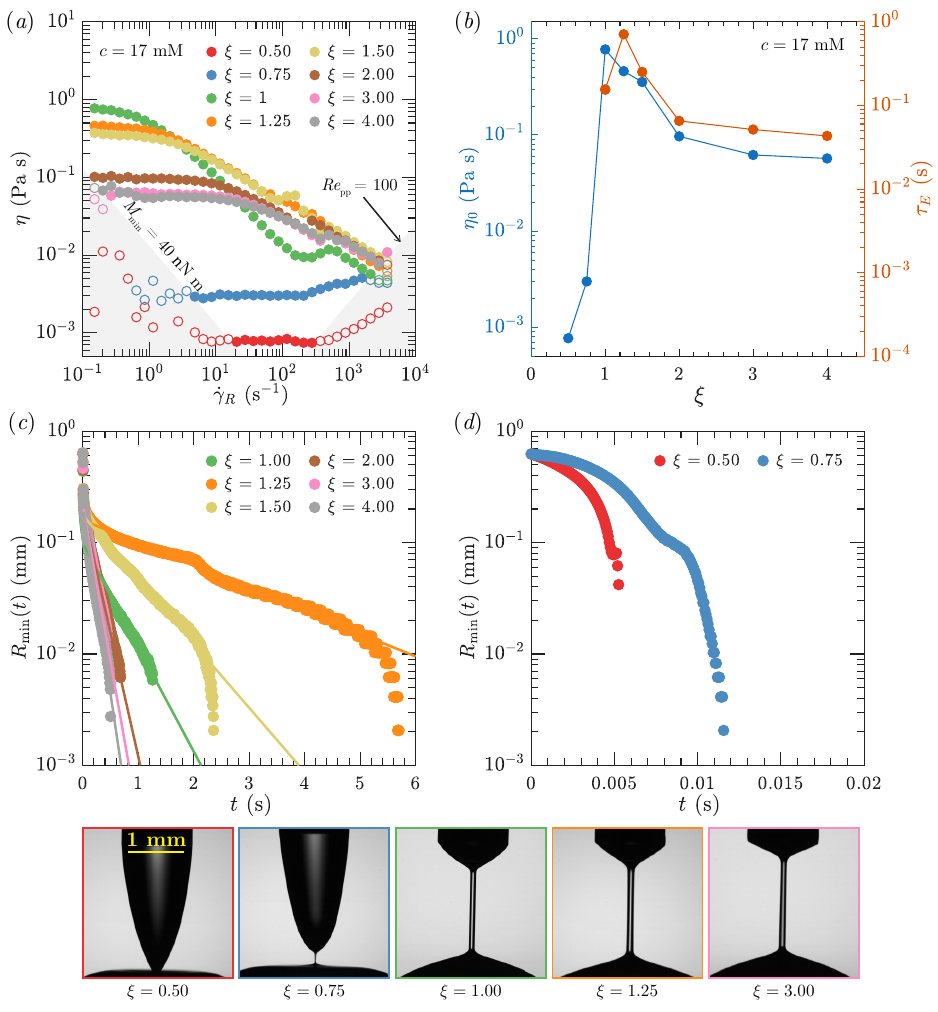}
\caption{Measurements of (\textit{a}) the steady shear viscosity $\eta(\dot{\gamma}_R)$ for a solution with a TTAC concentration of $c = 17~\mathrm{mM}$ and different levels of counter-ion (sodium salicylate with a molar ratio of $\xi = c_{ci}/c$). The leftmost grey shaded region in (\textit{a}) is the low torque limit of the rheometer $M_{\mathrm{min}}\leq40~$\SI{}{\nano\newton}~m. The rightmost grey shaded region in (\textit{a}) is the limit above which inertial instabilities become important, as represented by the Reynolds number $Re_{\mathrm{pp}}\geq100$. Measurements below and above these limits, respectively, are represented with hollow symbols. The blue left axis in (\textit{b}) shows the zero-shear--rate viscosity determined from (\textit{a}) as a function of the molar ratio, $\xi$. The righthand (orange) ordinate axis in (\textit{b}) shows measurements of the extensional relaxation time, $\tau_E$, from dripping-onto-substrate (DoS) extensional rheometry. The time-varying minimum filament radius $R_{\mathrm{min}}(t)$ measured from DoS rheometry for the non-dilute wormlike micelle solutions with $\xi\geq1.0$ are shown in (\textit{c}). Thin solid coloured lines represent fits of the elasto-capillary regime using the theory of \citet{Entov1997} represented by \eqnref{eqn:EC}. Measurements of $R_{\mathrm{min}}(t)$ for the less viscous surfactant solutions with lower counter-ion concentrations, $\xi = 0.5$ and $\xi = 0.75$, are shown separately in (\textit{d}) for clarity. Representative images of the filament during break-up are shown in the bottom panels of the figure.}
\label{ion}
\end{figure*}

In \figref{ion} we show measurements of the shear and extensional rheology of surfactant solutions with a constant TTAC concentration of $c=17~\mathrm{mM}$, but varying degrees of sodium salicylate, such that the molar ratio of counter-ion to surfactant is within the range of $0.5\leq\xi\leq4.0$. Small volumes (20 ml) of each test fluid were prepared for each molar ratio, $\xi$, by combining and diluting the concentrated master solutions of TTAC ($c = 69~\mathrm{mM}$) and sodium salicylate ($c_{ci} = 156~\mathrm{mM}$) with deionized water. Because of the very low weight fraction of TTAC and NaSal used in formulating these solutions, we take $\rho = 998~\mathrm{kg~m^{-3}}$ as the density of all test fluids. \figref{ion}(\textit{a}) shows the steady shear viscosity as a function of rim shear rate in a parallel plate geometry $\eta(\dot{\gamma}_R)$ for the TTAC solutions with different counter-ion ratios, $\xi$, measured using a parallel plate geometry with a plate radius of $R = 12.5~\mathrm{mm}$ and a gap of $d = 0.5~\mathrm{mm}$ attached to a torsional rheometer (MCR 703, Anton Paar). Given that the torsional flow between the parallel plates is an inhomogeneous shear flow, the rim shear rate $\dot{\gamma}_R = \varOmega R/d$ is used to define the maximum deformation rate at a given condition, where $\varOmega$ is the angular velocity of the upper plate. A correction must be used to adjust the measured shear stress (and the computed viscosity) for the non-homogeneous distribution of stress across the plate for shear-thinning fluids \citep{Bird1987}. The viscosity measurements become noisy below a torque of $M_{\mathrm{min}} = 40~\mathrm{nN~m}$. Above a Reynolds number of $Re_{\mathrm{pp}} = \rho\dot{\gamma}_R d^2/\eta \approx 100$ \citep{Davies2008}, inertial instabilities cause the reported viscosity value to artificially increase as shown, for example, by the red markers close to $\dot{\gamma}_R = 300~\mathrm{s}^{-1}$ (corresponding to the TTAC solution with the lowest counter-ion content of $\xi = 0.5$ and a viscosity similar to water, $\eta_s = 0.89~\mathrm{mPa~s}$). The grey shaded regions in \figref{ion}(\textit{a}) represent the limits of low torque $M_{\mathrm{min}}\leq40~\mathrm{nN~m}$ and inertial instabilities $Re_{\mathrm{pp}}\geq100$, beyond which the viscosity cannot be reliably measured. 

As the counter-ion ratio increases from $\xi = 0.5$ to $\xi = 1.0$, the viscosity of the TTAC solutions generally increase. While the solution with the lowest ratio, $\xi = 0.5$, has a viscosity similar to water, the TTAC solution with $\xi=1.0$ (as shown by the green symbols in \figref{ion}(\textit{a})), has a zero-shear--rate viscosity that is nearly 1000 times larger than the solvent ($\eta_0 = 0.80~\mathrm{Pa~s}$). The shear rheogram of the most viscous surfactant solution (with a molar ratio of $\xi=1.0$), exhibits a well-defined plateau and a constant zero-shear--rate viscosity at low shear rates ($\dot{\gamma}_R\ll0.1\mathrm{~s}^{-1}$), but a high degree of shear thinning at shear rates in excess of $\dot{\gamma}_R>1~\mathrm{s}^{-1}$. At a shear rate of $\dot{\gamma}_R = 100~\mathrm{s}^{-1}$ the viscosity of the TTAC solution begins to increase, but at a shear rate corresponding to Reynolds number significantly below the values needed for onset of inertial instabilities (corresponding to $Re_{\mathrm{pp}}=100$). This is due to a flow instability, which will be discussed and characterized in greater detail in \S~\ref{sec:FIB} using flow-induced birefringence. However, for now, we point out that generally all surfactant solutions with a counter-ion to surfactant molar ratio of $\xi\geq0.75$ exhibit the same augmented viscosity at shear rates of $60 \leq \dot{\gamma}_R \leq 200~\mathrm{s}^{-1}$ that are less than the threshold for inertial instabilities. The shear-thinning viscosity of the TTAC solutions with molar ratios of $\xi\geq1.0$ are fitted with the model proposed by \citet{Carreau1972}

\begin{equation}
    \eta(\dot{\gamma}_R) = \eta_0\Big(1+(\dot{\gamma}_R/\dot{\gamma}_0)^2 \Big)^{-m/2}
    \label{eqn:Carreau}
\end{equation}

\noindent
where $\dot{\gamma}_0$ is the shear rate above which the viscosity transitions from a constant zero-shear--rate viscosity $\eta_0$ to shear thinning, and $m$ is the power law index.

As the counter-ion ratio is increased beyond $\xi>1$, the zero-shear--rate viscosity starts to decrease again. The nonmonotonic trend in the viscosity with increasing counter-ion concentration is best revealed by the leftmost ordinate axis of \figref{ion}(\textit{b}) (coloured blue), which shows the zero-shear--rate viscosity as a function of the molar ratio of counter-ion to surfactant. An over-saturation of counter-ion and the reduction in viscosity  is caused by branching of the micelles \citep{Cui2007,Chellamuthu2008}, and the data suggests that the molar ratio of $\xi\approx1.0$ leading to the largest zero-shear--rate viscosity produces the longest wormlike micelles at equilibrium. Measurements of the extensional relaxation time, $\tau_E$, shown on the right axis (coloured orange) in \figref{ion}(\textit{b}), also exhibit a nonmonotonic trend with increasing counter-ion ratios.

The extensional relaxation times of the $c = 17~\mathrm{mM}$ TTAC solutions were measured using dripping-onto-substrate (DoS) capillarity-driven extensional rheometry \citep{Dinic2015}. This method relies on the formation of an unstable liquid filament, that thins under the action of surface tension. To produce this liquid filament, a pendant drop is formed at the end of a blunt-tip needle with an outer radius of $R_0 = 0.64~\mathrm{mm}$. A partially wetting substrate (made of a polyvinyl chloride sheet) is slowly brought into contact with the underside of the pendant drop, which is positioned approximately $6R_0$ from the tip of the needle. Rapid lateral spreading of the droplet drives the formation of the long, slender liquid bridge \citep{Zinelis2024}. The radius of the thin liquid filament then decays continuously over time until break-up, due to the action of capillary forces and the Rayleigh-Plateau instability. The time-evolving filament radius is measured from backlit images collected using a high-speed camera (NOVA P16, Photron Inc.). For a more comprehensive review of DoS extensional rheometry, and a more detailed description of the experimental methods, we direct readers to our prior dedicated investigation of DoS \citep{Warwaruk2026a}. 

Measurements of the time-evolving minimum filament radius, $R_{\mathrm{min}}(t)$, for solutions with a TTAC concentration of $c = 17~\mathrm{mM}$ and counter-ion molar ratios of $\xi\geq1.0$, are shown in \figref{ion}(\textit{c}). Generally, all solutions shown in \figref{ion}(\textit{c}) exhibit a region of exponential filament thinning of the form

\begin{equation}
    R_{\mathrm{min}}(t) = R^*\exp\Big(-\frac{(t-t^*)}{3\tau_E} \Big)
    \label{eqn:EC}
\end{equation}

\noindent
where $t^*$ and $R^*$ are the time and radius at which the flow transitions to \textit{elasto-capillary} thinning, and $\tau_E$ is the apparent extensional relaxation time \citep{Entov1997}. An exception is the solution with a molar ratio of $\xi = 1.25$ (shown using orange symbols) which exhibits exponential thinning, but also a peculiar local ``dip'' in values of $R_{\mathrm{min}}(t)$ at time $t = 2~\mathrm{s}$. A video of the time-evolving filament for the $\xi = 1.25$ solution is attached as a Supplemental Movie (\url{https://doi.org/10.6084/m9.figshare.34919400}), and reveals a wave that travels along the axial direction of the slender filament from the needle tip to the lower droplet. It is unclear what causes the traveling wave; \citet{Recktenwald2019} demonstrated that solutions of the surfactant cetyltrimethylammonium chloride or CTAC (with $n=16$ carbon atoms in its hydrophobic alkyl tail) mixed with the counter-ion sodium salicylate exhibit a banding instability when subjected to capillarity-driven extensional flow. It is plausible that axial asymmetry of the filament shape (inherent in DoS rheometry) results in gradients in surfactant concentration along the liquid-vapour interface that drive a Marangoni flow. That said, this peculiarity is unique to the $\xi = 1.25$ solution (near the local maximum in viscosity) and does not appear in any of the other test fluids shown in \figref{ion}(\textit{c}). Given this local maximum in the viscosity, stress-induced compositional variations that lead to local separation into two phases (with different local concentrations but similar viscosities) might thus develop. Since extensional rheometry is not the primary focus of the present investigation, we only briefly comment on this intriguing phenomenon, and defer deeper study for more dedicated future investigations on capillarity-driven extensional rheometry of wormlike micelle solutions. 

Despite this oddity, fits using \eqnref{eqn:EC} can still be performed on the measurements shown in \figref{ion}(\textit{c}) for the solution with a molar ratio of $\xi=1.25$, but we limit these to the second exponential thinning regime (at times $t>2~\mathrm{s}$).

Fits to the elastocapillary regime using the theory of \citet{Entov1997}, represented by \eqnref{eqn:EC}, are performed on all measurements shown in \figref{ion}(\textit{c}) of the time-varying minimum filament radius for solutions with molar ratios of $\xi\geq1.0$ and are represented by the solid coloured lines. Within the fitted regimes, the filaments were thin and slender (along the axial direction) as represented by the sample images shown at the bottom of \figref{ion}. Regression to the predictions of \eqnref{eqn:EC} were not performed on TTAC solutions with lower counter-ion ratios, i.e., $\xi = 0.5$ and 0.75. These solutions are characterized by a very short time-to-breakup, no apparent regime of exponential thinning, and a strongly axially varying filament profile, as shown respectively by the measurements of their time-varying filament radius in \figref{ion}(\textit{d}) and the representative sample images shown at the bottom of \figref{ion}. 

The non-dilute wormlike micelles solutions, with a counter-ion to surfactant molar ratio of $\xi\geq1.0$, exhibit large extensional relaxation times between $40 \leq\tau_E\leq 800~\mathrm{ms}$, as shown in \figref{ion}(\textit{b}). The measured values of the extensional relaxation time also decrease as a function of the molar ratio for $\xi>1.25$, consistent with our measurements for the zero-shear--rate viscosity $\eta_0(\xi)$. Therefore, branching of micelles with an excess of counter-ions produces both lower viscosities and lower extensional relaxation times.

\subsection{Microstructure and Linear Viscoelasticity}

Based on \S~\ref{sec:ion} it was shown that a counter-ion to surfactant molar ratio of $\xi = 1$ (for a TTAC concentration of $c = 17~\mathrm{mM}$) produced the largest viscosity with increasing molar ratio, $\xi$. Here we examine the linear viscoelastic rheology and microstructure of a $c = 34~\mathrm{mM}$ ($10~\mathrm{g~l^{-1}}$) TTAC solution with the same $\xi=1$ molar ratio. The higher viscosity of this more concentrated solution helps facilitate better signal contrast for visualizing the microstructure using cryogenic transmission electron microscope (cryo-TEM) imaging.

\begin{figure*}[h!]
\centering
\includegraphics[width=\textwidth]{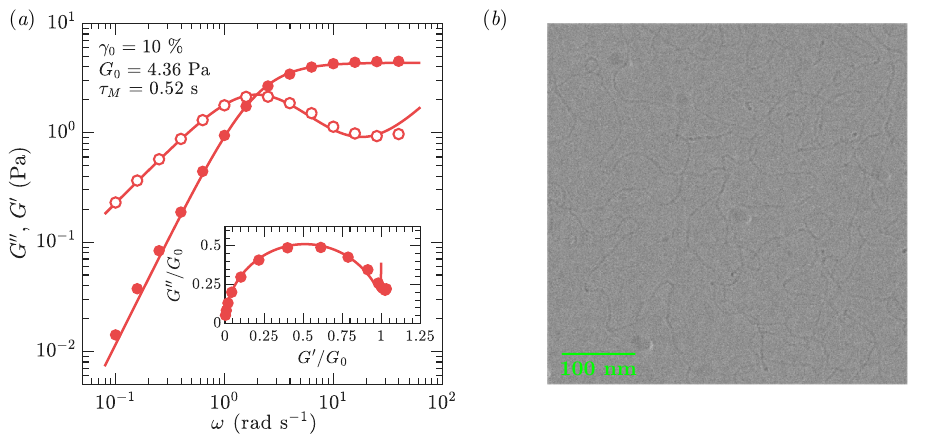}
\caption{The linear viscoelasticity of a solution of TTAC with a concentration of $c = 34~\mathrm{mM}$ and molar ratio of $\xi = 1$ is shown in (\textit{a}). Open red symbols in (\textit{a}) represent the loss modulus $G^{\prime\prime}(\omega)$, while filled red symbols correspond to the storage modulus $G^{\prime}(\omega)$. The strain amplitude was maintained constant across all frequencies at $\gamma_0 = 10~\%$. The red solid line in (\textit{a}) represents a fit to the measurements using the Jeffreys model described by \eqnref{eqn:UCM} with a plateau modulus of $G_0 = 4.36~\mathrm{Pa}$, a Maxwell relaxation time of $\tau_M = 0.52~\mathrm{s}$ and a high-frequency viscosity of $\eta_{\infty} = 0.025~\mathrm{Pa~s}$. The inset axis in (\textit{a}) shows a Cole-Cole plot of $G^{\prime\prime}/G_0$ versus $G^{\prime}/G_0$ for the measurements (shown with solid red symbols) and the modified Maxwell model of \eqnref{eqn:UCM}, represented with a solid red line. The image shown in (\textit{b}) is a micrograph taken using cryo-TEM, showing the wormlike micelles formed under these solution conditions ($c=34~\mathrm{mM}$, $\xi = 1$).}
\label{LVE}
\end{figure*}

Measurements of linear viscoelasticity were performed using an $R = 21.5~\mathrm{mm}$ parallel plate geometry (MCR 703, Anton Paar) with a gap of $d = 0.5~\mathrm{mm}$. Measurements were performed at a constant strain amplitude of $\gamma_0 = 10~\%$ and decreasing angular frequency from $\omega = 40~\mathrm{rad~s^{-1}}$ to $\omega = 0.1~\mathrm{rad~s^{-1}}$. \figref{LVE}(\textit{a}) shows the resulting storage modulus $G^{\prime}(\omega)$ and loss modulus $G^{\prime \prime}(\omega)$, as a function of the upper plate angular frequency $\omega$, for the TTAC solution. Non-dilute solutions of wormlike micelles are a rare example of a viscoelastic fluid that are well described by a single-mode Maxwell model \citep{Larson1999}. Here, the measurements are fit to a single-mode Jeffreys model (which also accounts approximately for the subsequent increase in $G^{\prime\prime}(\omega)$ at higher frequencies), i.e.,

\begin{subequations}\label{eqn:UCM}
  \begin{align}
    G^{\prime}(\omega) & = \frac{G_0(\tau_M\omega)^2}{1+(\tau_M\omega)^2}, \label{eqn:Gp} \\
    G^{\prime\prime}(\omega) & = \frac{G_0\tau_M\omega}{1+(\tau_M\omega)^2}+\eta_{\infty}\omega, \label{eqn:Gpp}
  \end{align}
\end{subequations}

\noindent
where $G_0$ is the plateau modulus, $\tau_M$ is the Maxwell relaxation time, and $\eta_{\infty}$ is the high-frequency viscosity. \eqnref{eqn:UCM} was fit to the measurements of storage and loss moduli shown in \figref{LVE}(\textit{a}) for the $c = 34~\mathrm{mM}$ TTAC solution with a molar ratio of $\xi = 1$, resulting in a plateau modulus of $G_0 = 4.36~\mathrm{Pa}$, a relaxation time of $\tau_M = 0.52~\mathrm{s}$ and a high-frequency viscosity of $\eta_{\infty} = 0.025~\mathrm{Pa~s}$. Agreement between the experimental measurements of $G^{\prime\prime}$ and $G^{\prime}$ with the adapted Jeffreys model of \eqnref{eqn:UCM} is also demonstrated by the Cole-Cole plot shown in the inset figure of \figref{LVE}(\textit{a}).

The average mesh size of the overlapping wormlike micelle solution can be estimated using rubber elasticity theory according to

\begin{equation}
    \zeta = \Big(\frac{k_BT}{G_0}\Big)^{1/3},
    \label{eqn:RET}
\end{equation}

\noindent
where $k_B$ is the Boltzmann constant, and $T=298~\mathrm{K}$ is the rheological test temperature \citep{Helgeson2010}. Based on the fitted plateau modulus of $G_0 = 4.36~\mathrm{Pa}$, the average mesh size of the overlapping micelles is estimated to be $\zeta = 98~\mathrm{nm}$.

Cryo-TEM images of the wormlike micelles formed in the $c = 34~\mathrm{mM}$ solution of TTAC with a molar ratio of $\xi = 1$ are shown in \figref{LVE}(\textit{b}). A 3~\SI{}{\micro\litre} aliquot of the TTAC solution was applied to lacey carbon grids (300~mesh Cu, Ted Pella, Inc.) that were glow-discharged at -15~mA for 30~s using a glow discharger (Emitech K100$\times$, Quorum Technologies). A vitrification device (Vitrobot Mark~IV, Thermo Fisher Scientific) was used to first remove excess sample from the grids by blotting for 6 to 12 s, and then immediately plunge-freeze the samples in liquid ethane. Images were collected using a 100~kV Tundra Cryo-TEM equipped with a Falcon C direct electron detector (Thermo Fisher Scientific) at a nominal magnification of $110~000\times$, corresponding to a pixel size of 2.4~\SI{}{\angstrom}~pixel$^{-1}$. The micelles in \figref{LVE}(\textit{b}) visibly overlap in several regions.  These overlaps show a mesh spacing that is qualitatively consistent with the value $\zeta = 98~\mathrm{nm}$ predicted from rubber elasticity theory of \eqnref{eqn:RET}.

\subsection{Flow-induced Birefringence}\label{sec:FIB}

\begin{figure*}[h!]
\centering
\includegraphics[width=\textwidth]{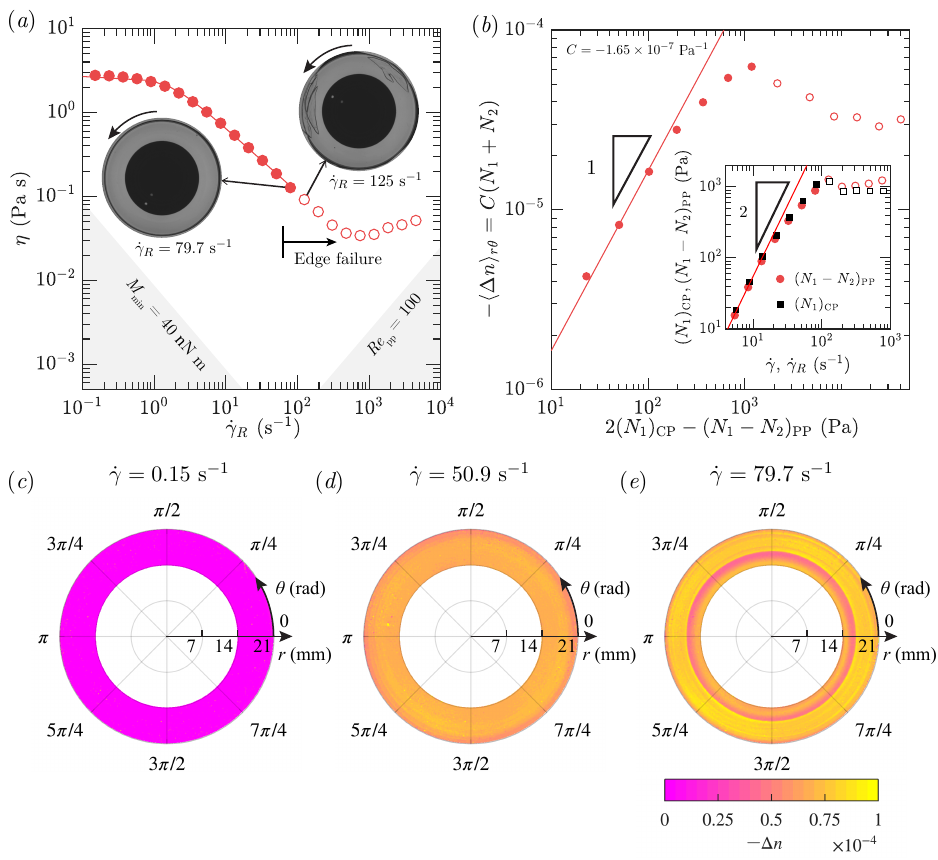}
\caption{Measurements of (\textit{a}) steady shear viscosity $\eta(\dot{\gamma})$ and (\textit{b}) the negative spatially averaged flow-induced birefringence $-\langle\Delta n\rangle_{r,\theta}$ as a function of normal stress difference $N_1+N_2\approx2(N_1)_{\mathrm{CP}}-(N_1-N_2)_{\mathrm{PP}}$, for the $c = 34~\mathrm{mM}$ TTAC solution with a molar ratio of $\xi = 1$ and $\tau_M = 0.52~\mathrm{s}$. The measurements in (\textit{a}) are collected using the transparent Rheo-Iris system, consisting of a parallel plate geometry with a glass upper plate of radius $R = 21.5~\mathrm{mm}$ and a gap height of $d = 0.5~\mathrm{mm}$. The solid red line in (\textit{a}) represents the fitted Carreau model of \eqnref{eqn:Carreau} with a zero-shear--rate viscosity of $\eta_0 = 2.7~\mathrm{Pa~s}$, a critical shear rate of $\dot{\gamma}_0 = 1.6~\mathrm{s}^{-1}$ and a power-law index of $m = 0.76$. Sample monochromatic images taken using the Rheo-Iris camera are shown in (\textit{a}) for two shear rates of $\dot{\gamma} = 79.7~\mathrm{s}^{-1}$ and $\dot{\gamma} = 125~\mathrm{s}^{-1}$, before and after edge failure, respectively. Open symbols in (\textit{a}) and (\textit{b}) denote measurements corrupted by edge failure. The solid red line in (\textit{b}) represents linear proportionality between birefringence and normal stress difference, i.e., $\langle\Delta n\rangle_{r,\theta} \propto N_1+N_2$, with a slope equal to the stress-optical coefficient $C$, as per \eqnref{eqn:C}. The inset axis in (\textit{b}) shows measurements of the first normal stress difference $(N_1)_{\mathrm{CP}}$ as a function of shear rate $\dot{\gamma}$ collected using the cone-and-plate geometry (black symbols) and also the normal stress difference $(N_1-N_2)_{\mathrm{PP}}$ as a function of the rim shear rate $\dot{\gamma}_R$ measured using the parallel-plate geometry of the Rheo-Iris system (red markers). The solid red line in the inset of (\textit{b}) shows the normal stress difference scaling for a second-order fluid $N_1 \propto \dot{\gamma}^2$. Polar axes (\textit{c}), (\textit{d}) and (\textit{e}) represent contours of birefringence $\Delta n$ for the flows shown in (\textit{a}) at shear rates of $\dot{\gamma}_R = 0.15~\mathrm{s}^{-1}$, $50.9~\mathrm{s}^{-1}$ and $79.7~\mathrm{s}^{-1}$, respectively.}
\label{FIB}
\end{figure*}

Steady shear viscosity measurements of the $c = 34~\mathrm{mM}$ solution of TTAC with a molar ratio of $\xi = 1$, previously shown in \figref{LVE}, are presented in \figref{FIB}(\textit{a}). Measurements of shear viscosity $\eta(\dot{\gamma})$ are collected using the same $R = 21.5~\mathrm{mm}$ parallel plate geometry with a gap of $d = 0.5~\mathrm{mm}$. The upper and lower plates of the geometry are optically transparent and made of quartz glass, permitting simultaneous optical flow measurements, in this case flow-induced birefringence (FIB). The torsional rheometer (MCR 703, Anton Paar) is equipped with a rheo-polarizing imaging setup, coined Rheo-Iris, previously detailed in \citet{araida2025} and \citet{Yamagata2026}.

The Rheo-Iris imaging setup consists of a ring-shaped LED (mounted to the upper spindle head of the rheometer) that projects circularly polarized light through the sample and onto the sensor of a high-speed polarization camera (CRYSTA PI-5WP; Photonic Lattice Inc.) that is mounted below the lower, stationary plate. The polarization camera consists of a 0.5~Mpixel sensor, where $2\times2~\mathrm{pixel}^2$ partitions of the sensor are overlaid with linear polarizers having four different orientations (each pixel polarization state differing by $45^{\circ}$ with its neighbour). Differences in the light intensity of the neighbouring pixels (each with a unique polarization state) are used to establish the local retardation $\delta$ and orientation $\phi$ \citep{Kusuno2025}. Sample monochromatic images taken using the Rheo-Iris system are shown in \figref{FIB}(\textit{a}) at two different shear rates $\dot{\gamma}_R$. The stem that mounts the upper plate to the rheometer spindle head obstructs measurements for radial positions of $0 \leq r \leq 14~\mathrm{mm}$, producing the distinct ``halo'' images shown in \figref{FIB}(\textit{a}).

\begin{figure*}[h!]
\centering
\includegraphics[width=\textwidth]{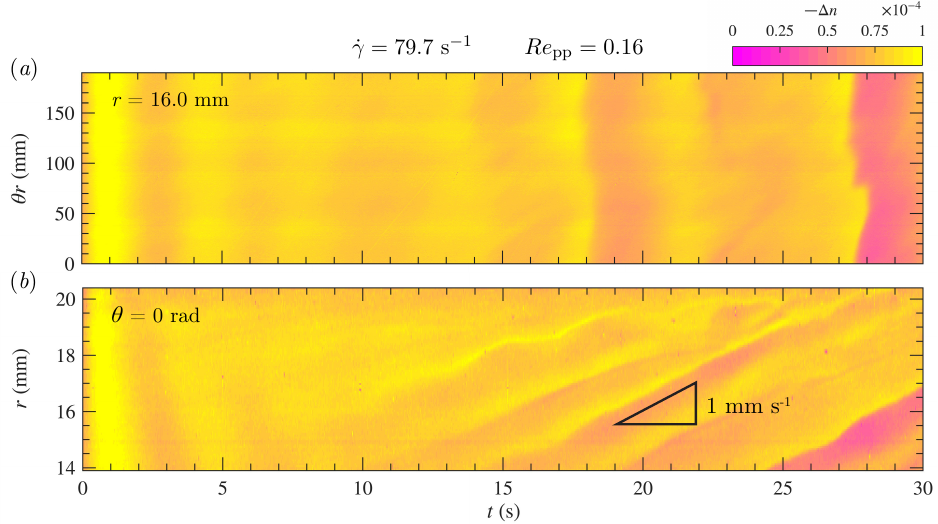}
\caption{Kymographs showing spatio-temporal variations in birefringence $\Delta n$ for the $c = 34~\mathrm{mM}$ TTAC solution with a molar ratio of $\xi = 1$ (same composition as used in \figref{FIB}) at a shear rate of $\dot{\gamma}_R = 79.7~\mathrm{s}^{-1}$, a Weissenberg number of $Wi_R = 41.4$, and a Reynolds number of $Re_{\mathrm{pp}} = 0.16$. Kymographs are shown for (\textit{a}) different angular positions $\theta$ (at a fixed radial position of $r = 16.0~\mathrm{mm}$), and (\textit{b}) different radial position $r$ (at a fixed angular position $\theta=0~\mathrm{rad}$). The snapshot in \figref{FIB}(\textit{e}) corresponds to the same measurements shown in the present figure, but at a time of $t = 27~\mathrm{s}$. In the Supplemental Movie (\url{https://doi.org/10.6084/m9.figshare.34919451}) we show an animation of the time-evolving birefringence and the banding instability for this flow.}
\label{Kymo}
\end{figure*}

Polarized light travels along the gradient direction of the flow, and therefore, the retardation measured by the polarization camera can be related to a projection of the refractive index tensor onto the imaging plane (aligned in the streamwise-vorticity plane or 1-3 plane), i.e.,

\begin{equation}
    \delta = (\Delta n)d = (n_{11}-n_{33})d
    \label{eqn:FIB}
\end{equation}

\noindent
where $\Delta n$ is the flow-induced birefringence, while $n_{11}$ and $n_{33}$ are components of the refractive index tensor acting along the streamwise direction and vorticity direction. Therefore, by rearranging \eqnref{eqn:FIB}, the birefringence can be assessed based on the measured retardation according to $\Delta n=\delta/d$. Furthermore, the birefringence can be related to stress using the \textit{stress-optical law}

\begin{equation}
    \mathbf{n} = C\boldsymbol{\sigma},
\end{equation}

\noindent
where $\boldsymbol{\sigma}$ is the total stress tensor, $\mathbf{n}$ is the refractive index tensor, and $C$ is the stress optical coefficient (with units of $\mathrm{Pa}^{-1}$). Given the configuration of the present imaging setup, the measured birefringence is a function of a projection of the stress tensor:

\begin{equation}
    \Delta n = n_{11}-n_{33} = C(\sigma_{11}-\sigma_{33})=C(N_1+N_2),
    \label{eqn:ntoN}
\end{equation}

\noindent
where $N_1 = \sigma_{11}-\sigma_{22}$ is the first normal stress difference, and $N_2 = \sigma_{22}-\sigma_{33}$ is the second normal stress difference. For wormlike micellar solutions, such as TTAC and CTAC, the stress optical coefficient is negative ($C<0$), resulting in a higher refractive index along the vorticity direction of the flow in comparison to the flow-direction, i.e., $n_{11}<n_{33}$ \citep{Ober2011}. 

The normal stress difference measured by the parallel plate rheometer is denoted $(N_1-N_2)_{\mathrm{PP}}$, where the subscript ``PP'' denotes the parallel plate geometry, inherent to the Rheo-Iris setup. Additional shear viscosity measurements of the same TTAC solutions were performed using a cone-and-plate geometry, with a radius of $R = 20~\mathrm{mm}$ and cone angle of $\alpha=1^{\circ}$, in order to measure the first normal stress difference $(N_1)_{\mathrm{CP}}$ (where the subscript ``CP'' denotes cone-and-plate). By contrasting the normal stress measurements collected using the parallel plate and cone-and-plate configurations at the same shear rate we can calculate the combination of normal stress differences $N_1+N_2= 2(N_1)_{\mathrm{CP}}-(N_1-N_2)_{\mathrm{PP}}$. We can then determine the stress optical coefficient of the test fluid by rearranging \eqnref{eqn:ntoN}, i.e.,

\begin{equation}
    C =  \frac{\langle\Delta n \rangle_{r \theta}}{ 2(N_1)_{\mathrm{CP}}-(N_1-N_2)_{\mathrm{PP}}}
    \label{eqn:C}
\end{equation}

\noindent
where $\langle \Delta n\rangle_{r,\theta}$ is the spatially-averaged birefringence for all measured radial positions ($14\leq r \leq 21.5~\mathrm{mm}$) and angular positions ($0\leq \theta \leq 2\pi$) of the parallel plate.

From \figref{FIB}(\textit{a}) it is clear that the $c = 34~\mathrm{mM}$ TTAC solution with a molar ratio of $\xi = 1$ exhibits a shear-thinning viscosity that is well described by the Carreau model of \eqnref{eqn:Carreau} with a zero-shear--rate viscosity of $\eta_0 = 2.7~\mathrm{Pa~s}$, a critical shear rate of $\dot{\gamma}_0 = 1.6~\mathrm{s}^{-1}$ and a power-law index of $m = 0.76$ \citep{Carreau1972}. From monochromatic images using the Rheo-Iris system, edge failure and the entrainment of bubbles into the sample is observed at shear rates of $\dot{\gamma}_R \geq 125~\mathrm{s}^{-1}$, denoted using hollow markers in \figref{FIB}. Although edge fracture is clear from images of the flow at $\dot{\gamma}_R \geq 125~\mathrm{s}^{-1}$, measurements of the viscosity appear to be a poor indicator of edge fracture and do not exhibit a noticeable change in value until much larger rates of $\dot{\gamma}_R \approx 800~\mathrm{s}^{-1}$. Although not shown, measurements of the shear viscosity collected using the cone-and-plate geometry $\eta_{\mathrm{CP}}(\dot{\gamma})$ for the same fluids overlap with the parallel-plate shear rheogram shown in \figref{FIB}(\textit{a}).

The evolution in the spatially-averaged flow-induced birefringence $\langle \Delta n\rangle_{r,\theta}$ is shown in \figref{FIB}(\textit{b}), and plotted with respect to the normal stress difference, $N_1+N_2\approx 2(N_1)_{\mathrm{CP}}-(N_1-N_2)_{\mathrm{PP}}$, measured using the parallel-plate and cone-and-plate geometries. The inset axis in \figref{FIB}(\textit{b}) shows the normal stress differences of the TTAC solution measured using the parallel-plate $(N_1-N_2)_{\mathrm{PP}}$ and cone-and-plate $(N_1)_{\mathrm{CP}}$ geometries, as a function of shear rate. At low shear rates of $\dot{\gamma} \lesssim 20~\mathrm{s}^{-1}$, the normal stress differences exhibit consistency with a second-order fluid where $(N_1)_{\mathrm{CP}}\propto\dot{\gamma}^2$ and $(N_1-N_2)_{\mathrm{PP}} \propto \dot{\gamma}_R^2$. Furthermore, the magnitude of the normal stress differences measured using the different geometries are similar, implying the second normal stress difference $N_2$ is small in comparison to the first normal stress difference, and consequently cannot be reliably resolved from the current measurements. 

Our measurements with the Rheo-Iris system show that at low magnitudes of the normal stress difference, $(N_1+N_2) \lesssim 200~\mathrm{Pa}$, the average birefringence scales proportionally with the normal stresses, $-\langle \Delta n\rangle_{r,\theta} \propto (N_1+N_2)$, consistent with the stress-optical law of \eqnref{eqn:ntoN} and represented by the solid red line in \figref{FIB}(\textit{b}). The slope of the linear function of $-\langle \Delta n\rangle_{r,\theta}$ versus $N_1+N_2$ produces a stress optical coefficient of $C = -1.65\times10^{-7}~\mathrm{Pa^{-1}}$ that is comparable in magnitude to previously reported values of similar surfactant systems (e.g., CTAC) \citep{Ober2011}. 

At higher rates, the normal stress differences shown in the inset of \figref{FIB}(\textit{b}) decrease slightly in magnitude near the critical rate at which an instability develops and edge fracture occurs, i.e., $\dot{\gamma}_R \geq 125~\mathrm{s}^{-1}$, equivalent to Weissenberg numbers of $Wi_R = \tau_M\dot{\gamma}_R\geq65$. This corresponds to the large reduction in the magnitude of the average birefringence, $-\langle \Delta n\rangle_{r,\theta}$, at normal stresses of $(N_1-N_2)_{\mathrm{PP}} > 1000~\mathrm{Pa}$. Therefore, although the flow-induced instability and edge fracture cannot be easily detected from measurements of the viscosity, cf. \figref{FIB}(\textit{a}), measurements of the normal stresses provide a more sensitive probe of the formation features that disrupt viscometric flow.

Instantaneous snapshots of the birefringence, $\Delta n(r,\theta)$, at shear rates of $\dot{\gamma}_R = 0.15~\mathrm{s}^{-1}$, $50.9~\mathrm{s}^{-1}$ and $79.7~\mathrm{s}^{-1}$ are shown in \figref{FIB}(\textit{c}), (\textit{d}), and (\textit{e}), respectively. For the low and moderate shear rates of $\dot{\gamma}_R = 0.15~\mathrm{s}^{-1}$, and $50.9~\mathrm{s}^{-1}$, shown in \figref{FIB}(\textit{c}) and (\textit{d}), the birefringence appears spatially homogeneous for the range of radial coordinates measured using the Rheo-Iris system. However, at a shear rate of $\dot{\gamma}_R = 79.7~\mathrm{s}^{-1}$, just before edge failure, the flow becomes unstable with bands of birefringence that vary along the radial (vorticity) direction. Kymographs that show the spatio-temporal variation of birefringence following onset of instability in the TTAC solution, are presented in \figref{Kymo}. We highlight that this instability develops under conditions that are seemingly devoid of inertia, with a Reynolds number of $Re_{\mathrm{pp}} = 0.16$, but a large Weissenberg number of $Wi_R  =41.4$. The kymograph of azimuthal birefringence, $\Delta n(\theta,t)$, for fixed radial position of $r = 16.0~\mathrm{mm}$, but different angular positions is shown in \figref{Kymo}(\textit{a}). This shows clear bands of birefringence that are uniform along the azimuthal direction -- consistent with the instantaneous snapshot shown in \figref{FIB}(\textit{e}) (corresponding to time $t = 27~\mathrm{s}$). A kymograph of radial birefringence, $\Delta n(r,t)$, at a fixed angular position of $\theta = 0~\mathrm{rad}$, but different radial positions is shown in \figref{Kymo}(\textit{b}). The bands of differing birefringence advect radially outward at a speed of approximately $1~\mathrm{mm~s^{-1}}$ or about $2.5~\%$ of the rim velocity of the upper plate, i.e., $0.025\dot{\gamma}_Rd$. The structure of the flow immediately preceding the onset of flow instability and edge fracture, appears to exhibit vorticity banding, characterized by rings of constant birefringence that differ along the radial (vorticity) direction. A video of the time-evolving vorticity bands is shown in the attached Supplemental Movie (\url{https://doi.org/10.6084/m9.figshare.34919451}).

\section{Dilute Solutions of Wormlike Micelles}

In the previous section, it was shown that non-dilute surfactant solutions exhibit a range of non-Newtonian properties including: viscoelasticity (consistent with a single-mode Jeffreys model), shear-thinning, normal stresses, flow-induced birefringence and banding instabilities (both in shear and extension). However, these solutions are too viscous to be relevant for turbulent drag reduction applications. Instead, dilute surfactant solutions, with a viscosity comparable to the solvent, are desired as drag-reducing fluids. Here we consider the same TTAC surfactant, but at significantly lower concentrations than those considered in \S~\ref{sec:semi-dilute}. We prepare a larger volume (300~ml) of a $c = 3.4~\mathrm{mM}$ ($1~\mathrm{g~l^{-1}}$) TTAC solution, by diluting the $c=69~\mathrm{mM}$ TTAC master solution discussed in \S~\ref{sec:test_fluids}. Two relative counter-ion concentrations were considered. The first was a counter-ion molar ratio of $\xi=1$, as this produced the largest viscosity for a fixed TTAC concentration in \S~\ref{sec:semi-dilute} (albeit for a larger overall TTAC concentration). The second condition was an over-saturated counter-ion molar ratio of $\xi=5$. The additional counter-ion was incorporated by adding and diluting the necessary amount of the $c_{ci} = 156~\mathrm{mM}$ solution of sodium salicylate, discussed in \S~\ref{sec:test_fluids}. The ability of these fluids to reduce turbulent drag is examined in a bespoke turbulent Taylor-Couette flow facility.

\subsection{Flow-induced Structures}\label{sec:SIS}

\begin{figure*}[h!]
\centering
\includegraphics[width=\textwidth]{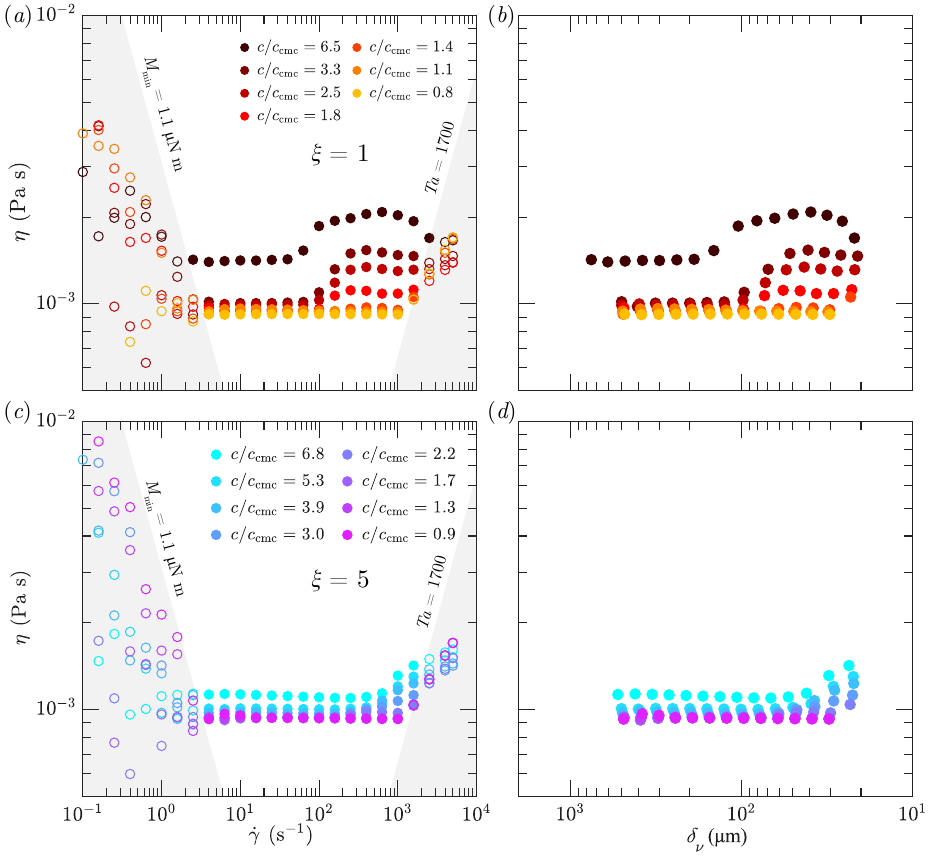}
\caption{Measurements of steady shear viscosity $\eta$ shown with respect to (\textit{a}) shear rate $\dot{\gamma}$ and (\textit{b}) the viscous length scale $\delta_{\nu} = \eta_0/\sqrt{\rho\sigma_w}$, for dilute TTAC solutions with a counter-ion molar ratio of $\xi = 1$ and a critical micelle concentration of $c_{\mathrm{cmc}}=0.53~\mathrm{mM}$. Similar measurements of (\textit{c}) $\eta(\dot{\gamma})$ and (\textit{d}) $\eta(\delta_{\nu})$, are shown for dilute TTAC solutions with a larger counter-ion molar ratio of $\xi = 5$, and a critical micelle concentration of $c_{\mathrm{cmc}} = 0.25~\mathrm{mM}$. The direction of the abscissa in (\textit{b}) and (\textit{d}) is reversed to parallel the appearance of (\textit{a}) and (\textit{c}). Grey shaded regions in plots of $\eta(\dot{\gamma})$ shown on axes (\textit{a}) and (\textit{c}) correspond to the limits of low torque sensitivity $M_{\min} = 1.1$~\SI{}{\micro\newton}~m and onset of centripetal flow instabilities, at which the Taylor number exceeds $Ta = 1700$. Hollow symbols are measurements that fall below the torque sensitivity limit, or above the threshold for centripetal flow instabilities.}
\label{SIS}
\end{figure*}

The steady shear viscosity $\eta(\dot{\gamma})$ of the dilute solutions of TTAC were measured using a torsional rheometer (DHR-3, TA Instruments) equipped with a double-gap concentric cylinder fixture. The double gap concentric cylinder geometry has a larger wetted surface area than the parallel plate geometries used in \S~\ref{sec:semi-dilute}, which improves the accuracy of viscosity measurements that are close in value to the solvent viscosity of water, $\eta \rightarrow \eta_s = 0.89~\mathrm{mPa~s}$. The double-gap geometry consists of a stationary annular cup with an inner radius of $R_1 = 19.9~\mathrm{mm}$ and an outer radius of $R_2 = 22.4~\mathrm{mm}$. The rotor, mounted to the spindle head of the rheometer, has an inside radius of $R_3 = 20.4~\mathrm{mm}$ and an outside radius of $R_4 = 22.0 \mathrm{mm}$. Therefore, the average rotor radius is $R = (R_3+R_4)/2 = 21.2~\mathrm{mm}$, and the average gap is $d = \big((R_3-R_1)+(R_2-R_4)\big)/2=0.45~\mathrm{mm}$. Fluid is loaded between the gaps of the concentric cylinder with an immersion height $L = 59.0~\mathrm{mm}$. This geometry can be used for rheometry provided the flow established is a homogeneous cylindrical Couette flow. However, centripetal forces drive the formation of Taylor vortex instabilities in the narrow gap for a Taylor number of $Ta \geq 1700$, where the Taylor number is defined as $Ta = \rho^2\varOmega^2Rd^3/\eta^2=Re^2(d/R)$, the angular velocity of the rotor is $\varOmega = \dot{\gamma}d/R$, and the Reynolds number is $Re = \rho\varOmega R d/\eta$ \citep{Ewoldt2015}.

The shear viscosity as a function of increasing shear rate $\eta(\dot{\gamma})$ for several dilute solutions of TTAC with a counter-ion molar ratio of $\xi = 1$ is shown in \figref{SIS}(\textit{a}). Seven solutions are shown for concentrations ranging from $0.8\leq c/c_{\mathrm{cmc}}\leq 6.5$, where the critical micelle concentration for test fluids with a molar ratio of $\xi = 1$ is $c_{\mathrm{cmc}} = 0.53~\mathrm{mM}$ (cf. \figref{CMC}). Solutions were progressively diluted from the $3.4~\mathrm{mM}$ ($1~\mathrm{g~l^{-1}}$) stock solution in the turbulent Taylor-Couette flow facility, which will be discussed in greater detail in \S~\ref{sec:TTCF}. Small aliquots (7~ml) were extracted from the turbulent flow facility for the shear viscosity measurements shown in \figref{SIS}. Measurements at low shear rates are limited by the torque sensitivity of the rheometer, which we found to be $M_{\mathrm{min}} \approx 1.1$~\SI{}{\micro\newton}~m. Measurements are also ignored at larger shear rates when the Taylor number exceeds $Ta = 1700$ and the flow becomes non-viscometric. At shear rates of $\dot{\gamma} \lesssim 100~\mathrm{s}^{-1}$, the viscosity of the TTAC solutions are constant and independent of shear rate. The zero-shear--rate viscosity $\eta_0$ increases slightly with concentration $c/c_{\mathrm{cmc}}$. At a concentration of $c/c_{\mathrm{cmc}}=6.5$ the zero-shear--rate viscosity is $\eta_0 = 1.6\eta_s$ and shows a greater enhancement compared to the lower concentration solutions -- suggesting that the growing micelles begin to overlap under equilibrium conditions. At a critical shear rate of $\dot{\gamma}_c \approx 100~\mathrm{s}^{-1}$ the viscosity of the TTAC solutions with $c/c_{\mathrm{cmc}} \geq 1.8$ exhibit an apparent shear thickening, consistent with prior observations of flow-induced structure (FIS) formation in similar cationic surfactant solutions \citep{Ohlendorf1986,Bewersdorff1988}. The magnitude of the shear-induced enhancement in viscosity increases with TTAC concentration. At concentrations below the critical micelle concentration $c<c_{\mathrm{cmc}}$, the viscosity exhibits no signatures of flow-induced structures and remains constant across all measured shear rates.

In \figref{SIS}(\textit{b}) we present the same measurements of the shear viscosity as \figref{SIS}(\textit{a}), but replotted with respect to a representative viscous length scale of the flow $\delta_{\nu} = \eta_0/\sqrt{\rho \sigma_w}$, where $\sigma_w$ is the wall shear stress measured by the rheometer. This is, perhaps, an unconventional way of presenting steady shear viscosity measurements. However, the viscous length scale is an important parameter in wall-bounded turbulence, and is used to help identify the characteristic length scale of viscous effects in relation to inertia, i.e., the friction Reynolds number $Re^* = d/\delta_{\nu}$. This will be discussed in greater detail in \S~\ref{sec:TTCF} when we focus on turbulent Taylor-Couette flow. As the shear rate and stress increase, the viscous length scale decreases $\delta_{\nu} \propto \sigma_w^{-1/2}$. Therefore to preserve and parallel the appearance of the shear rheograms $\eta(\dot{\gamma})$ shown in \figref{SIS}(\textit{a}), the direction of the abscissa in \figref{SIS}(\textit{b}) is reversed, instead showing the viscosity with decreasing values of the viscous length. The critical viscous length scale for the formation of flow-induced structures at $\xi =1$ can be identified to be $ \delta_{\nu,c} = 100~$\SI{}{\micro\meter}.

Steady shear viscosity measurements of $\eta(\dot{\gamma})$ are also shown in \figref{SIS}(\textit{c}) for TTAC solutions prepared at a higher salicylate counter-ion ratio, $\xi = 5$. Measurements are shown for eight solutions with TTAC concentrations over the range $0.9\leq c/c_{\mathrm{cmc}}\leq6.8$, where the critical micelle concentration is $c_{\mathrm{cmc}} = 0.25~\mathrm{mM}$. Similar to the TTAC solutions at the lower molar ratio shown in \figref{SIS}(\textit{a}), the zero-shear--rate viscosity of the $\xi=5$ TTAC solutions shown in \figref{SIS}(\textit{c}) appears to monotonically increase with TTAC concentration. A subtle augmentation in the viscosity (indicative of FIS formation) occurs at a critical shear rate of $\dot{\gamma}_c\approx 400~\mathrm{s}^{-1}$ that precedes the development of Taylor vortex instabilities. This critical shear rate is larger than the value for the lower counter-ion ratio shown in \figref{SIS}(\textit{a}). The flow-induced structures of the $\xi = 5$ solutions also appear to produce less of an augmentation in viscosity; however, the influence of inertia and the formation of Taylor vortices obscure the observation of the local maximum in the measured viscosity that can be seen for the $\xi = 1$ TTAC solutions in \figref{SIS}(\textit{a}). When plotted with respect to the viscous length scale, cf. \figref{SIS}(\textit{d}), the critical shear rate for the formation of flow-induced structures corresponds to a viscous length scale of $ \delta_{\nu,c} = 50~$\SI{}{\micro\meter}.

\begin{figure*}[h!]
\centering
\includegraphics[width=\textwidth]{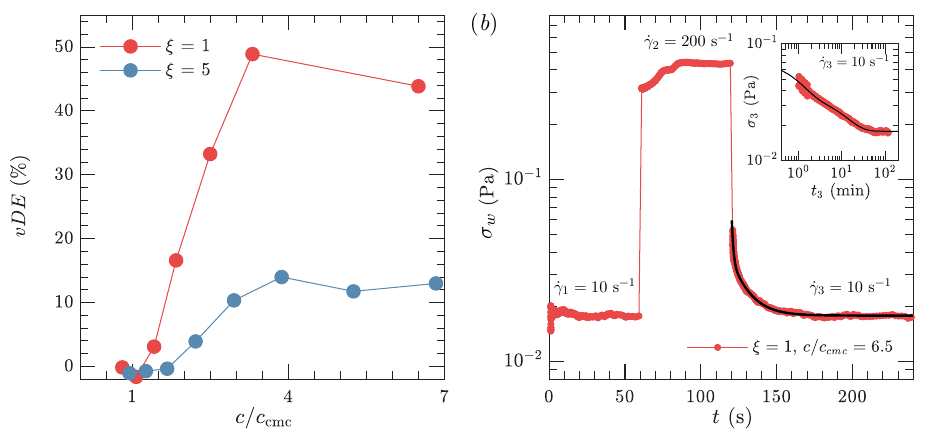}
\caption{(\textit{a}) Maximum viscous drag enhancement, attributed to the increase in the shear viscosity caused by flow-induced structures, defined using \eqnref{eqn:vDE} using $\delta_\nu = \delta_{\mathrm{FIS}} = 50~$\SI{}{\micro\meter} for the $\xi=1$ TTAC solutions and $\delta_{\mathrm{FIS}} = 33~$\SI{}{\micro\meter} for the $\xi = 5$ TTAC solutions. Plots of (\textit{b}) the time-varying shear stress acquired from a three-interval stress growth experiment for the TTAC solution with a concentration of $c/c_{\mathrm{cmc}} = 6.5$ and a counter-ion molar ratio of $\xi = 1$. The three interval test consists of impulsive step changes in the shear rate, the first at $\dot{\gamma}_1=10~\mathrm{s}^{-1}$, the second at a higher rate of $\dot{\gamma}_2=200~\mathrm{s}^{-1}$, and finally a return to the lower shear rate of $\dot{\gamma}_3=10~\mathrm{s}^{-1}$. The inset axis in (\textit{b}) shows the time-decaying stress measured over the third interval (at $\dot{\gamma}_3=10~\mathrm{s}^{-1}$) on double logarithmic axes. The black solid line in (\textit{b}), and the inset axis, represent the double exponential function of \eqnref{eqn:3iTT} with a stress plateau of $\sigma_{\infty} = 0.018~\mathrm{Pa}$, a relaxation time of $\tau_f = 0.97~\mathrm{s}$, a break-up time of $\tau_b = 10.9~\mathrm{s}$ and viscosities of $\eta_f = 3.68~\mathrm{mPa~s}$ and $\eta_b = 1.81~\mathrm{mPa~s}$.}
\label{3iTT}
\end{figure*}

The data in \figref{SIS} shows that flow-induced structures produce an augmented viscosity at relatively large shear rates. To describe the relative change in the viscosity caused by flow-induced structures, we define a \textit{viscous drag enhancement} as

\begin{equation}
    vDE(\delta_{\nu}) = \Big(\frac{\eta(\delta_{\nu})-\eta_0}{\eta_0}\Big) \times 100~\%
    \label{eqn:vDE}
\end{equation}

\noindent
where the viscous drag enhancement depends on the shear rate, cf. \figref{SIS}(\textit{a,~c}), but also, equivalently, on the viscous length scale $\delta_{\nu}$, which can more readily be related to conditions within the turbulent Taylor-Couette flow. \figref{3iTT}(\textit{a}) shows the viscous drag enhancement plotted with respect to TTAC concentration. Here the viscous drag enhancement is assessed for a given viscous length scale of $\delta_{\mathrm{FIS}} = 50~$\SI{}{\micro\meter} (equivalent to a wall shear stress of $0.34\lesssim\sigma_w\lesssim0.81~\mathrm{Pa}$, depending on TTAC concentration and $\eta_0$) for the $\xi = 1$ TTAC solution -- corresponding approximately to the maximum enhancement in viscosity shown in \figref{SIS}(\textit{b}). For the $\xi =5$ TTAC solution, the viscous drag enhancement of \eqnref{eqn:vDE} is evaluated using a viscous length scale of $\delta_{\mathrm{FIS}} = 33~$\SI{}{\micro\meter} (equivalent to a wall shear stress of $0.81\lesssim\sigma_w\lesssim1.14~\mathrm{Pa}$, depending on TTAC concentration and $\eta_0$). \figref{3iTT}(\textit{a}) shows that the degree of viscosity enhancement caused by FIS formation is greater for the $\xi = 1$ TTAC solution than for solutions with larger amounts of salicylate ($\xi = 5$). The maximum enhancement in viscosity caused by flow-induced structures is $vDE = 49.5~\%$ for the TTAC solution with concentration of $c/c_{\mathrm{cmc}} = 3.3$ and a counter-ion molar ratio of $\xi = 1$.

To evaluate the time scale of FIS development, a three-interval stress growth experiment, shown in \figref{3iTT}(\textit{b}), is performed on the TTAC solution with a concentration of $c/c_{\mathrm{cmc}} = 6.5$ and a counter-ion molar ratio of $\xi = 1$. The critical shear rate for the inception of flow-induced structures is $\dot{\gamma}_c = 50~\mathrm{s}^{-1}$, as shown by the dark red markers in \figref{SIS}(\textit{a}). Therefore, the three-interval stress growth experiment consists of step changes in the imposed shear rate, commencing below the onset of FIS formation ($\dot{\gamma}_1=10~\mathrm{s}^{-1} = 0.2\dot{\gamma}_c$), next increasing to conditions above FIS inception ($\dot{\gamma}_2=200~\mathrm{s}^{-1} = 4\dot{\gamma}_c$), and finally a return to a shear rate of $\dot{\gamma}_3=10~\mathrm{s}^{-1} = 0.2\dot{\gamma}_c$. From inspection of \figref{3iTT}(\textit{b}), it is clear that significant time is needed to develop and destroy the flow-induced structures when transitioning from the shear rate of $\dot{\gamma}_1 = 10~\mathrm{s}^{-1}$ to $\dot{\gamma}_2 = 200~\mathrm{s}^{-1}$, and vice versa (i.e., from $\dot{\gamma}_2$ to $\dot{\gamma}_3$). The sample is maintained at $\dot{\gamma}_1$ for $\Delta t = 1~\mathrm{min}$, then $\dot{\gamma}_2$ for $\Delta t = 1~\mathrm{min}$, and lastly $\dot{\gamma}_3$ for $\Delta t = 2~\mathrm{min}$. The resulting stress as a function of time, $\sigma_w(t)$, is shown for the three intervals in \figref{3iTT}(\textit{b}). The inset axis shows the decay in the stress as a function of time for the third interval, i.e., $\sigma_3(t_3)$, where $t_3 = t-2~\mathrm{min}$ is the time that starts at the transition from step 2 to 3.

Similar stress growth experiments have been performed in previous works to characterize the time scale of FIS build-up and decay in dilute surfactant solutions \citep{Ohlendorf1986,Zheng2000}. \citet{Ohlendorf1986} used a stress relaxation approach to measure the time scale for FIS decay of a similar $c = 3.4~\mathrm{mM}$ TTAC solution with an equimolar mixture of sodium salicylate ($\xi=1$) as that shown in \figref{3iTT}(\textit{b}); albeit, their tests were performed at a lower temperature of $T = 293~\mathrm{K}$. Using a single-gap Couette geometry ($d = 0.32~\mathrm{mm}$), \citet{Ohlendorf1986} found that the critical shear rate for forming flow-induced structures was $\dot{\gamma}_c = 40~\mathrm{s}^{-1}$ -- not dissimilar to the critical shear rate of $\dot{\gamma}_c = 50~\mathrm{s}^{-1}$ determined for the present test fluid shown in \figref{SIS}(\textit{a}). \citet{Ohlendorf1986} found that measurements of stress relaxation for self-assembling surfactant solutions of TTAC/NaSal followed a double exponential trend, where the shortest time scale $\tau_{f}$ was denoted an inherent relaxation time of the flow-induced structures, as it did not vary with the rate of FIS formation (i.e., with $\dot{\gamma}_2$). On the other hand, the longer time scale $\tau_b$, a representative time scale for break-up of the flow-induced structures, was found to depend strongly on the rate of formation. \citet{Ohlendorf1986} measured the relaxation time of the flow-induced structures to be short and generally $\tau_{f}\leq \mathcal{O}(1~\mathrm{ms})$, while the break-up time varied widely for the same TTAC solution, $1<\tau_{b} <1~000 ~\mathrm{s}$, depending on the shear rate $\dot{\gamma}_2$ used to form the supramolecular microstructure. To describe the stress relaxation $\sigma_3(t_3)$ of the test fluid shown in \figref{3iTT}(\textit{b}), a double exponential function, similar to that used in \citet{Ohlendorf1986}, was used:

\begin{equation}
    \sigma_3(t_3)-\sigma_{\infty} = \eta_{f} \dot{\gamma}_3 \exp\Big( -\frac{t_3}{\tau_{f}}\Big)+ \eta_{b} \dot{\gamma}_3 \exp\Big(-\frac{t_3}{\tau_{b}} \Big) 
    \label{eqn:3iTT}
\end{equation}

\noindent
where $\sigma_{\infty}$ is the plateau stress at large times, while $\eta_{f}$ and $\eta_b$ are the changes in viscosity caused by relaxation and break-up of the flow-induced structures, respectively. The fitted double exponential function of \eqnref{eqn:3iTT} is represented by the black solid line in \figref{3iTT}(\textit{b}). Fits to the time-varying stress $\sigma_3(t_3)$ using \eqnref{eqn:3iTT} produce a stress plateau of $\sigma_\infty = 0.018~\mathrm{Pa}$ (identical to the stress prior to FIS formation $\sigma_1 = 0.018~\mathrm{Pa}$). Therefore, after a sufficient amount of time $t_3\gg\tau_b$, the stress plateaus to its equilibrium level, $\sigma_\infty=\eta_0\dot{\gamma}_3$. The relaxation time of the flow-induced structures was determined to be $\tau_{f} = 0.97~\mathrm{s}$, and the break-up time $\tau_b = 10.9~\mathrm{s}$ -- similar to the time scales measured by \citet{Ohlendorf1986} for a similar TTAC solution. The representative viscosities from fits to $\sigma_3(t_3)$ using \eqnref{eqn:3iTT} are $\eta_f = 3.68~\mathrm{mPa~s}$ and $\eta_b = 1.81~\mathrm{mPa~s}$. At short times $t_3\rightarrow0~\mathrm{s}$, the cumulative stress is $\sigma_3(0) = \sigma_\infty+\eta_{f}\dot{\gamma}_3+\eta_{b}\dot{\gamma}_3 = 0.073~\mathrm{Pa}$, and the viscosity is $\eta_3(0) = \sigma_3(0)/\dot{\gamma}_3 = 7.28~\mathrm{mPa~s}$, which is larger than the steady-state viscosity in the second interval at high rates $\eta_2 = \sigma_2/\dot{\gamma}_2 = 2.18~\mathrm{mPa~s}$. Therefore, although the stress decreases when transitioning from interval two to three, as shown in \figref{3iTT}(\textit{b}), the instantaneous viscosity can momentarily increase before decaying to its equilibrium value (i.e. $\eta_0$) as the structures break down. This suggests that rapid fluctuations in the instantaneous strain rate (common in turbulent flows) may also produce fluctuations in the local instantaneous viscosity in excess of the steady state values reported in \figref{SIS}.

\subsection{Turbulent Drag Modification}\label{sec:TTCF}

\begin{figure*}[h!]
\centering
\includegraphics[width=\textwidth]{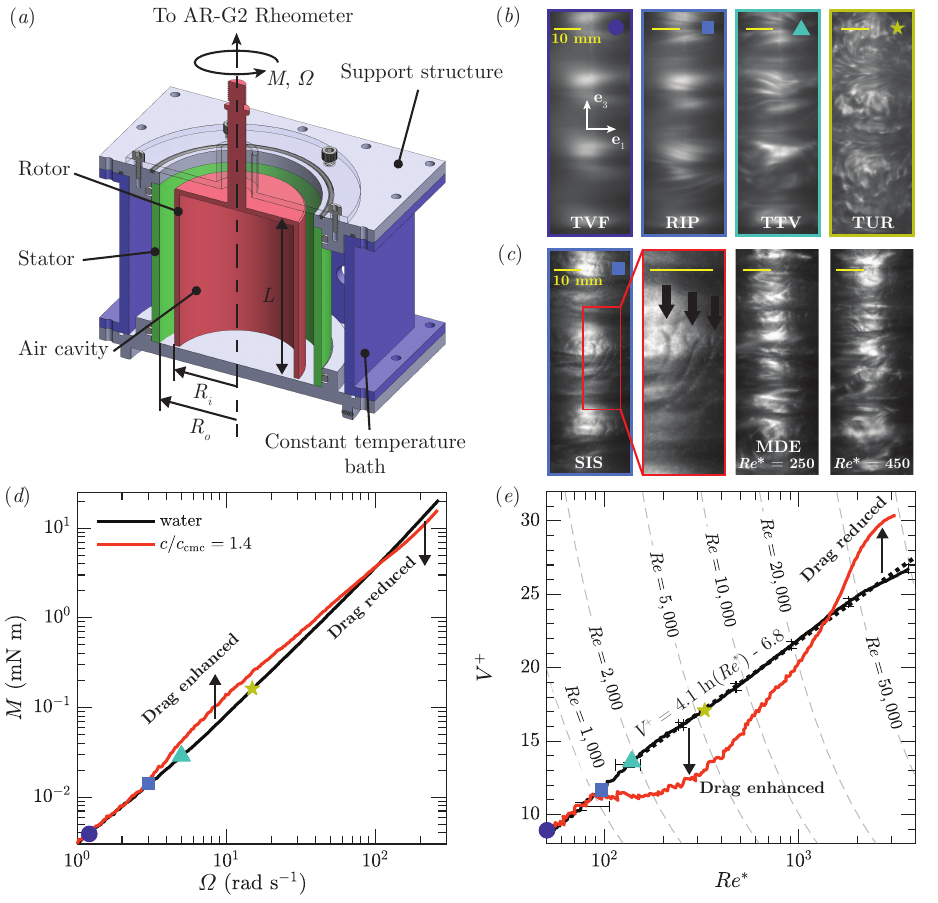}
\caption{A cross-section of a three-dimensional model of the wide-gap Taylor-Couette flow facility is shown in (\textit{a}). Kalliroscope flow visualization of water in different flow regimes is shown in (\textit{b}), which includes: Taylor vortex flow (TVF), ripple (RIP) Taylor vortex flow, turbulent Taylor vortex (TTV) flow, and featureless turbulence (TUR). Kalliroscope flow visualization of a surfactant solution with a TTAC concentration of $c/c_{\mathrm{cmc}} = 1.4$ and a counter-ion molar ratio of $\xi = 1$ is shown in (\textit{c}). The first two images (from left to right) in (\textit{c}) show the flow immediately at the onset of flow-induced structure formation, when the angular velocity is $\varOmega = 3~\mathrm{rad~s^{-1}}$ and the friction Reynolds number is $Re^*\approx100$. The second image in (\textit{c}), bordered by the red line, shows a magnified image of three spanwise-oriented G\"{o}rtler flow motions (identified with black arrows) seen at the inception of FIS formation. The last two images in (\textit{c}) are images of the flow in the regime of drag enhancement. Dimensional drag curves of torque $M$ versus angular velocity for the Newtonian flow of water and the non-Newtonian flow of the surfactant solution over nearly four decades of imposed torque are presented in (\textit{d}). Lastly, (\textit{e}) shows the evolution of non-dimensional drag curves in Prandtl-von K\'arm\'an coordinates for water and the surfactant solution. The black dashed line represents fits of the measurements of $V^+(Re^*)$ using the logarithmic form of \eqnref{eqn:PvK} to the Newtonian flow of water in the TUR regime ($300 \leq Re^* \leq 3000$). Border colours of the images in (\textit{b}) and (\textit{c}) correspond to the flow conditions shown with coloured markers in (\textit{d}) and (\textit{e}). In the Supplemental Movie (\url{https://doi.org/10.6084/m9.figshare.34919487}) we show an animation of the Kalliroscope flow visualization for TTAC.}
\label{TTCF}
\end{figure*}

The ability for dilute TTAC solutions to modify turbulent drag is assessed using a bespoke wide-gap Taylor-Couette flow facility. A cross-section of a 3-D rendering of the Taylor-Couette flow facility is shown in \figref{TTCF}(\textit{a}). The same design has been used by our group in previous investigations of turbulent drag modification using biopolymers (e.g., mucins and mucilage), synthetic polymers (polyethylene oxide, polyacrylamide, polyisobutylene) and superhydrophobic surfaces \citep{Rajappan2019,Rajappan2019b,Rajappan2020,Warwaruk2026b,More2026}. Test fluid (approximately 285~ml) is loaded between the two concentric cylinders. The outer glass cylinder, shown in green in \figref{TTCF}(\textit{a}), is held fixed and has an internal wall radius of $R_o = 38.9~\mathrm{mm}$. The inner aluminum cylinder or rotor, with a wall radius of $R_i = 25.4~\mathrm{mm}$ and shown in red in \figref{TTCF}(\textit{a}), is fastened to the spindle head of a stress-controlled rheometer (AR-G2, TA Instruments). A torque $M$ is applied to the inner rotor and an optical encoder measures the angular velocity $\varOmega$. The sample gap is $d = R_o-R_i=13.5~\mathrm{mm}$. The length of the rotor is $L = 81.7~\mathrm{mm}$, which is much larger than the gap ($L = 6d$). Within the rotor body, a hollow recess traps a pocket of air that provides a shear-free interface at the bottom of the rotor, mitigating end effects. The stator assembly is maintained at a constant temperature of $T=298~\mathrm{K}$ using a thermal bath with water that is recirculated through a thermoelectric chiller (Thermocube, AMS Technologies AG). The walls of the thermal bath are transparent and made of acrylic, permitting qualitative visualization of coherent flow motions, as shown in \figref{TTCF}(\textit{b}) and (\textit{c}).

Flow visualization is only performed for two test fluids: water (images of which are shown in \figref{TTCF}(\textit{b})), and a TTAC solution with a concentration of $c/c_{\mathrm{cmc}}=1.4$ and a counter-ion molar ratio of $\xi = 1$ (images of which are shown in \figref{TTCF}(\textit{c})). Kalliroscope particles (mica flakes) are added to the fluid, while a high-speed camera (Miro M310, Vision Research Inc.) and a diffuse LED light source are used to record images of the resulting coherent flow motions. Snapshots of the local flow structure are captured along the mean streamwise-direction ($\mathbf{e}_1$) and axial or vorticity direction ($\mathbf{e}_3$), as annotated on the first image of \figref{TTCF}(\textit{b}). 

With increasing rotation rate $\varOmega$, or Reynolds number, Taylor–Couette flow undergoes a sequence of flow instabilities, transitioning from an initial laminar flow, to the appearance of centripetally-driven Taylor vortices, and eventually approaching featureless turbulence \citep{Andereck1986}. Measurements of the torque $M$ as a function of the rotational velocity of the rotor are shown in \figref{TTCF}(\textit{d}). For all test fluids, the angular velocity of the rotor is slowly increased from $\varOmega = 1~\mathrm{rad~s^{-1}}$ to $\varOmega = 270~\mathrm{rad~s^{-1}}$ over a duration of $11~\mathrm{min}$ at a slow (logarithmic) growth rate of $d \ln\varOmega/dt = 9\times10^{-3}~\mathrm{s}^{-1}$. Measurements of the Newtonian flow of water are represented with the black solid line, while the corresponding non-Newtonian flow of the TTAC solution is shown in red. Flow visualization images shown in \figref{TTCF}(\textit{b}), have border colours that correspond to the conditions represented by the coloured markers in \figref{TTCF}(\textit{d}). At a low rotation rate of $\varOmega = 1.1~\mathrm{rad~s^{-1}}$, the flow exhibits three stationary Taylor vortex pairs -- as shown by the first image of \figref{TTCF}(\textit{b}) bordered in dark blue. Adopting the naming convention of \citet{Andereck1986}, we refer to this flow regime as Taylor vortex flow (TVF). As the rotational rate is increased further to $\varOmega = 3~\mathrm{rad~s^{-1}}$, ripples in the Taylor vortices form, as shown in the second panel of \figref{TTCF}(\textit{b}) and denoted as RIP. These ripples then break down into turbulent Taylor vortices (TTV) at a rotation rate of $\varOmega = 5~\mathrm{rad~s^{-1}}$; a regime where Taylor vortices are superposed with additional small scale turbulent fluctuations or G\"ortler vortices. Lastly, for sufficiently large rotation rates, $\varOmega \geq 15~\mathrm{rad~s^{-1}}$, the Taylor-Couette flow exhibits featureless turbulence (TUR), where the visual appearance of Taylor vortices are no longer detectable -- as shown by the yellow-bordered image in \figref{TTCF}(\textit{b}).

For the surfactant solution shown by the red solid line in \figref{TTCF}(\textit{d}), measurements of the torque as a function of rotation rate $M(\varOmega)$ overlap with the Newtonian flow of water within the TVF regime ($1 < \varOmega < 3~\mathrm{rad~s^{-1}}$). However, close to the onset of ripples in TVF ($\varOmega~=~3~\mathrm{rad~s^{-1}}$), the measured torque of the surfactant solution increases and starts to deviate from the Newtonian flow with increasing rotation rate. This enhanced torque, or \textit{drag enhancement}, persists for a wide range of rotational velocities $3~\leq~\varOmega~\leq~90~\mathrm{rad~s^{-1}}$, spanning the regimes of RIP to TUR. However, at even larger rotation rates of $\varOmega > 90~\mathrm{rad~s^{-1}}$, the flow of the surfactant solution transitions to a state of \textit{drag reduction}, with a torque that is now lower than the Newtonian TUR flow at the same rotational velocity. At the onset of drag enhancement (when $\varOmega = 3~\mathrm{rad~s^{-1}}$), the surfactant solution exhibits local G\"{o}rtler-type flow motions that extend along the spanwise (axial) direction, as shown in the first image of \figref{TTCF}(\textit{c}). The second image in \figref{TTCF}(\textit{c}) shows a magnified view of three successive spanwise G\"{o}rtler flow motions, stacked sequentially along the streamwise flow direction. We demonstrate below that the onset of drag enhancement is coupled to the conditions at which flow-induced structures are formed in steady shear rheometry (cf. \figref{SIS}). In the Supplemental Movie (\url{https://doi.org/10.6084/m9.figshare.34919487}) we show an animation of the Kalliroscope flow visualization for the same TTAC solution, depicting the onset of the flow-induced structures, marked by the appearance of spanwise-oriented G\"{o}rtler-like flow motions. 

In the Taylor–Couette flow, the rotor wall velocity is $V = \varOmega R_i$ and the mean wall shear stress is $\sigma_w = M/2\pi R_i^2L$. The Reynolds number is $Re = \rho V d /\eta_0$, where the zero-shear--rate viscosity is assessed using the shear viscosity measurements of \figref{SIS}, and the skin-friction coefficient is $C_f = 2\sigma_w/\rho V^2$. For the Newtonian flow of the solvent, i.e., water, the dynamic viscosity is $\eta_0 = \eta_s = 0.89~\mathrm{mPa~s}$. 
In a wall-driven turbulent flow, a viscous scaling is commonly used to normalize the rotor velocity and wall shear stress. The relevant velocity and length scales used for this scaling are the friction velocity $u^* = \sqrt{\sigma_w/\rho}$ and viscous length scale $\delta_{\nu}=\nu_0/ u^*$, where the zero-shear--rate kinematic viscosity is $\nu_0=\eta_0/\rho$. Using these quantities a \textit{friction Reynolds number} can be defined as:

\begin{equation}
    Re^* = \frac{\rho u^*d}{\eta_0}=\frac{d}{\delta_\nu}.
    \label{eqn:Re_tau}
\end{equation}

\noindent
Taylor-Couette flow in the turbulent (TUR) regime exhibits a logarithmic dependence of the rotor velocity on wall shear stress (i.e., $V\propto\ln\sigma_w$) and when non-dimensionalised by these characteristic viscous scales this relationship can be written in the form

\begin{equation}
    V^+ = \mathcal{N}\ln(Re^*)+\mathcal{M}
    \label{eqn:PvK}
\end{equation}

\noindent
where $V^+=V/u^*$ is the dimensionless rotor velocity, and $\mathcal{N}$ and $\mathcal{M}$ are device-dependent constants \citep{Lathrop1992}, which in Taylor-Couette flow depend on the curvature of flow; here, the ratio of the rotor radii is $\mathcal{R} =  R_i/R_o = 0.65$, and the curvature of the flow is defined as $\mathcal{C} = d/R_i = 0.53$. Note that the dimensionless rotor velocity and friction Reynolds number are commonly denoted as the Prandtl–von K\'arm\'an skin-friction coordinates and can be equivalently expressed in terms of the skin-friction coefficient as $V^+ = \sqrt{2/C_f}$ and $Re^* = Re\sqrt{C_f/2}$ \citep{Lathrop1992}. The present Taylor-Couette flow facility is able to achieve bulk Reynolds numbers of $350 \lesssim Re \lesssim 100~000$ corresponding to friction Reynolds numbers of $50 \lesssim Re^* \lesssim 3500$ (depending on the test fluid).

Non-dimensional drag curves in Prandtl–von K\'arm\'an coordinates of $V^+(Re^*)$ are shown in \figref{TTCF}(\textit{e}) for the Newtonian Taylor-Couette flow of water and the non-Newtonian TTAC solution with a concentration of $c/c_{\mathrm{cmc}}=1.4$. Nomograms of constant bulk Reynolds number, $Re\equiv Re^*V^+$, are denoted using grey dashed lines. For the Newtonian flow of water, the flow transitions to a TUR regime (cf. \figref{TTCF}(\textit{b})) above a bulk Reynolds number of $Re = 5000$, or a friction Reynolds number of $Re^* = 300$. In the TUR flow regime ($300\leq Re^*\leq3000$) measurements are well described by the logarithmic functional form of \eqnref{eqn:PvK} with $\mathcal{N} = 4.1$ and $\mathcal{M} = -6.8$, shown by the black dashed line in \figref{TTCF}(\textit{e}). Compared to the Newtonian flow of water, the TTAC solution exhibits values of the dimensionless rotor velocity $V^+(Re^*)$ that are \textit{lower} than those measured for water between friction Reynolds numbers of $100\leq Re^*\leq1500$ (i.e., a drag enhancement). In this drag-enhanced regime, Kalliroscope images reveal that the TTAC solution produces narrow (in comparison to the scale of TTV structures) coherent flow motions along the spanwise direction superposed with additional small scale fluctuations. At larger friction Reynolds numbers of $1500<Re^*\leq3000$ the surfactant solution exhibits values of $V^+(Re^*)$ that are \textit{larger} in magnitude than water (i.e., drag reduction). If measurements are performed at a constant angular velocity, a reduction in the torque will augment values of the dimensionless rotor velocity since $V^+\propto M^{-1/2}$, but will also decrease the friction Reynolds number since $Re^*\propto M^{1/2}$. This is the approach used in the current Taylor-Couette flow experiments. However, if experiments are performed at a constant torque, the friction Reynolds number defined by \eqnref{eqn:Re_tau} remains unchanged and a reduction in fluid frictional drag produces an enhancement in angular velocity $V^+\propto \varOmega$ (i.e., an enhanced throughput). 

To quantify the relative differences in the flow resistance of the surfactant drag-reduced flows with the baseline conditions of water, a \textit{drag modification percentage} is defined as 

\begin{equation}
    \mathcal{D}(Re^*) = \Bigg[1-\frac{M_S}{M_N} \Bigg]\times 100~\% \equiv \Bigg[1-\Bigg(\frac{V^+_{N}}{V^+_{S}} \Bigg)^2 \Bigg]\times 100~\%
    \label{eqn:DM}
\end{equation}

\noindent
where $M_N$ and $V^+_N$ are the torque and dimensionless rotor velocity for the Newtonian flow of water, and $M_S$ and $V^+_S$ are the torque and dimensionless rotor velocity for the non-Newtonian surfactant flows. The drag modification percentage is determined by comparing the dimensionless rotor velocity of the surfactant flows with water at a constant friction Reynolds number. Values of the drag modification that are negative ($\mathcal{D} <0~\%$) correspond to an enhancement in drag, while positive values ($\mathcal{D} > 0~\%$) indicate drag reduction. 

\begin{figure*}[h!]
\centering
\includegraphics[width=\textwidth]{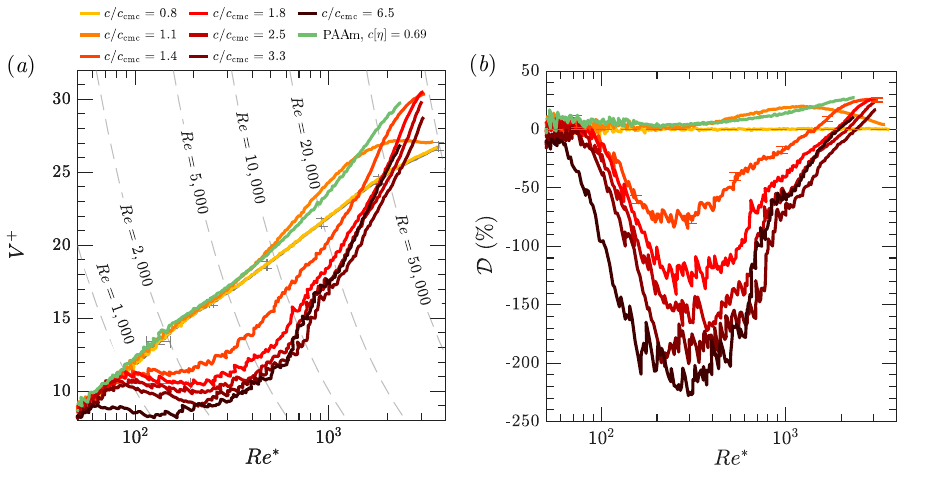}
\caption{Plots of (\textit{a}) Prandtl-von K\'arm\'an drag curves $V^+(Re^*)$ and (\textit{b}) the drag modification percentage, defined from \eqnref{eqn:DM}, for solutions of seven TTAC concentrations with a counter-ion molar ratio of $\xi = 1$. Recall the critical micelle concentration for this solution is $c_{\mathrm{cmc}} = 0.53~\mathrm{mM}$. Measurements of water are shown using a grey solid line in (\textit{a}) and a grey dashed line ($\mathcal{D} = 0~\%$) in (\textit{b}), but overlap with the TTAC solution below the critical micelle concentration $c/c_{\mathrm{cmc}}=0.8$. Measurements are compared to a drag-reducing solution of polyacrylamide PAAm, shown in green, at a concentration of $c[\eta] = 0.69$ (where the intrinsic viscosity of the solution is $[\eta] = 69.1~\mathrm{dl~g^{-1}}$). Grey dashed lines in (\textit{a}) represent nomograms of constant bulk Reynolds number $Re\equiv Re^*V^+$.}
\label{DR-1to1}
\end{figure*}

In \figref{DR-1to1}(\textit{a}) we present drag curves in Prandtl-von K\'arm\'an coordinates of TTAC solutions at seven different concentrations, all with a counter-ion molar ratio of $\xi = 1$. For each test fluid, the angular velocity is ramped up from $\varOmega = 1~\mathrm{rad~s^{-1}}$ to $\varOmega = 270~\mathrm{rad~s^{-1}}$ using the same procedure discussed in regards to \figref{TTCF}(\textit{c}). The first solution examined is the solution with the highest concentration of $c/c_{\mathrm{cmc}} = 6.5$. Upon completing a flow sweep, $70~\mathrm{ml}$ of test fluid is removed and replaced with deionized water, diluting the solution by 25~\%. The flow sweep is then repeated for each successive dilution. When removing solution from the Taylor-Couette flow facility, a small volume was retained for rheological testing, the results of which have been previously shown in \S~\ref{sec:SIS}. This post-shearing test procedure leverages the fact that wormlike surfactant solutions do not degrade when subjected to prolonged shearing. The seven TTAC concentrations span a range of values from slightly below the critical micelle concentration to well above: $0.8 \leq c/c_{\mathrm{cmc}} \leq 6.5$. For completeness and comparative purposes, measurements are also compared with analogous tests using an aqueous solution of polyacrylamide (PAAm) at a concentration of $c[\eta] = 0.69$ (where the intrinsic viscosity of the solution is $[\eta] = 69.1~\mathrm{dl~g^{-1}}$). For more details regarding the rheology and preparation of the PAAm solutions, consult \ref{app:PAAm}.

All solutions with a TTAC concentration of $c/c_\mathrm{cmc}\geq1.4$, exhibit drag enhancement at low friction Reynolds numbers (with values of $V^+$ less than those of water), and drag reduction (i.e., values of $V^+$ larger than water for a given $Re^*$) at higher friction Reynolds numbers. TTAC solutions with a concentrations of $1.4\leq c/c_\mathrm{cmc}\leq3.3$ exhibit a sudden transition to drag enhancement, at approximately $Re^* = 80$. Recall from \figref{SIS}(\textit{b}), that the same solutions exhibit flow-induced structures when the viscous length reaches a critical value of $\delta_{\nu,c} = 100~$\SI{}{\micro\meter}. This corresponds to a critical friction Reynolds number of $Re^*_c = d/\delta_{\nu,c} = 135$ -- similar in magnitude to the observed onset of drag enhancement $Re^* = 80$. The most concentrated TTAC solution with a concentration of $c/c_\mathrm{cmc}=6.5$ also exhibits a sudden transition to drag enhancement, but at a lower friction Reynolds number of $Re^* = 60$. This solution also exhibits a slightly larger critical viscous length scale for FIS formation, $\delta_{\nu,c} = 140~$\SI{}{\micro\meter}, indicating a slightly smaller critical friction Reynolds number of $Re_c^* = 95$. Note that unlike the laminar viscometric flows of \figref{SIS}, the Taylor-Couette flow depicted in \figref{DR-1to1} is spatio-temporally varying, even at these conditions of low Reynolds numbers -- corresponding to the TVF and RIP flow regimes depicted in \figref{TTCF}(\textit{b}). Therefore, it is likely that local fluctuations in the instantaneous shear rate may exceed the critical shear rate for FIS formation at conditions below the critical condition predicted by a viscometric flow, explaining the robustly consistent observation of an early onset to drag enhancement at $Re^* < Re^*_c$. To our knowledge, this sudden onset of drag enhancement has not been observed experimentally in Taylor-Couette flow of viscoelastic wormlike micellar solutions. Furthermore, this drag enhancement seems to be unique to wormlike micelles; our control experiments with the high molecular weight polymer PAAm tested here, do not exhibit drag enhancement for any concentration or at any Reynolds number (see \ref{app:PAAm} for more details).

\begin{figure*}[h!]
\centering
\includegraphics[width=\textwidth]{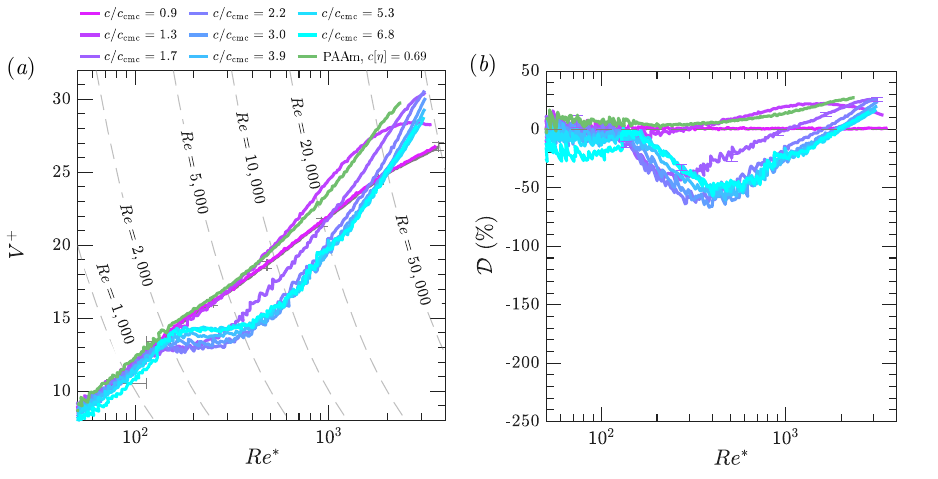}
\caption{Plots of (\textit{a}) Prandtl-von K\'arm\'an drag curves $V^+(Re^*)$ and (\textit{b}) the drag modification percentage, defined from \eqnref{eqn:DM}, for solutions of seven TTAC concentrations with a counter-ion molar ratio of $\xi = 5$. Recall the critical micelle concentration for this solution is $c_{\mathrm{cmc}} = 0.25~\mathrm{mM}$. Measurements of water are shown using a grey solid line in (\textit{a}) and a grey dashed line ($\mathcal{D} = 0~\%$) in (\textit{b}), but overlap with the TTAC solution below the critical micelle concentration $c/c_{\mathrm{cmc}}=0.9$. Measurements are compared to a drag-reducing solution of polyacrylamide PAAm, shown in green, at a concentration of $c[\eta] = 0.69$ (where the intrinsic viscosity of the solution is $[\eta] = 69.1~\mathrm{dl~g^{-1}}$). Grey dashed lines in (\textit{a}) represent nomograms of constant bulk Reynolds number $Re\equiv Re^*V^+$. Limits of the ordinate axis $-250\leq\mathcal{D}\leq50~\%$ are the same as \figref{DR-1to1}.}
\label{DR-1to5}
\end{figure*}

The drag modification percentage as a function of friction Reynolds number $\mathcal{D}(Re^*)$ is shown in \figref{DR-1to1}(\textit{b}) for the TTAC solutions with a counter-ion molar ratio of $\xi=1$. Within the regime of drag enhancement (corresponding to $\mathcal{D}<0~\%$), the level of drag modification generally becomes increasingly negative (i.e., more drag enhancement) with increasing TTAC concentration. The maximum drag enhancement (MDE), corresponding to the most negative value of the drag modification curve $\mathcal{D}(Re^*)$ occurs at a friction Reynolds number of $Re^*\approx300$ for all TTAC solutions with a concentration of $c/c_{\mathrm{cmc}}\geq1.4$. At higher friction Reynolds numbers, the TTAC solutions (with concentrations of $c/c_{\mathrm{cmc}}\geq1.4$) transition to a state of drag reduction (i.e., $\mathcal{D}>0~\%$). The friction Reynolds number at which the flow transitions from drag enhancement to drag reduction generally increases with TTAC concentration. At a friction Reynolds number of $Re^* = 3000$, the TTAC solution with a concentration of $c/c_{\mathrm{cmc}}=1.4$ achieves a drag reduction of $\mathcal{D} = 30~\%$ -- similar in magnitude to the values achieved using PAAm (shown by the green curve in \figref{DR-1to1}(\textit{b})), as well as other drag-reducing additives (e.g., polyethylene oxide, flax seed mucilage, and natural salivary mucins) used in previous investigations of turbulent drag reduction in a similar Taylor-Couette flow \citep{Rajappan2019,Rajappan2020,Warwaruk2026b}. At concentrations close to but still slightly larger than the critical micelle concentration ($c/c_{\mathrm{cmc}}=1.1$), the flow exhibits no drag enhancement, but still a sizable amount of drag reduction (as large as $\mathcal{D} = 20~\%$) for friction Reynolds numbers of $Re^*\geq300$. Reducing the concentration further, such that it falls slightly below the critical micelle concentration ($c/c_{\mathrm{cmc}}=0.8$) results in a drag curve that overlaps with water in \figref{DR-1to1}(\textit{a}) and no drag modification ($\mathcal{D}(Re^*) = 0~\%$) in \figref{DR-1to1}(\textit{b}). 

In \figref{DR-1to5}(\textit{a}) we present drag curves in Prandtl-von K\'arm\'an coordinates, $V^+(Re^*)$, for TTAC solutions at a larger counter-ion molar ratio ($\xi=5$) compared to those investigated in \figref{DR-1to1}. Similar to the results of \figref{DR-1to1}, TTAC solutions with a larger concentration of salicylate counter-ion (and concentrations of $1.7\leq c/c_{\mathrm{cmc}} \leq 6.8$) also exhibit an abrupt transition to a state of drag enhancement. The transition to drag enhancement occurs at a friction Reynolds number of approximately $Re^* = 150$, which is larger than the critical Reynolds number measured for the solutions with a lower counter-ion ratio of $\xi = 1$ (cf. \figref{DR-1to1}). The critical viscous length scale for FIS formation exhibited by the $\xi = 5$ TTAC solutions in \figref{SIS}(\textit{d}) was $\delta_{\nu,c} = 50~$\SI{}{\micro\meter}, implying flow-induced structures form at a friction Reynolds number of $Re_c^*=270$. This critical friction Reynolds number over-predicts the true onset of drag enhancement of $Re^*=150$ -- similar to the predictions for the TTAC solutions at a lower molar ratio of $\xi=1$ discussed above. However, bearing in mind the fluctuating nature of the Taylor-Couette flow states at these Reynolds numbers, it is still appropriate to connect the sudden onset of drag enhancement to shear-driven generation of flow-induced structures.

In \figref{DR-1to5}(\textit{b}) we replot the measurements of torque versus speed in terms of the drag modification percentage as a function of friction Reynolds number, $\mathcal{D}(Re^*)$, for the TTAC solutions with a counter-ion molar ratio of $\xi=5$. Solutions with concentrations of $1.7\leq c/c_{\mathrm{cmc}} \leq 6.8$, exhibit less overall drag enhancement than the TTAC solutions with a lower counter-ion molar ratio of $\xi=1$ shown in \figref{DR-1to1}(\textit{b}). The maximum drag enhancement (i.e., the most negative value of $\mathcal{D}$) is approximately $\mathcal{D} = -60~\%$ -- about four times lower in magnitude than the maximum drag enhancement $\mathcal{D} = -220~\%$ observed for the TTAC solution with a concentration of $c/c_{\mathrm{cmc}} = 6.5$ and counter-ion molar ratio of $\xi = 1$, shown in \figref{DR-1to1}(\textit{b}). However, at higher friction Reynolds numbers, the levels of drag reduction ($\mathcal{D}>0~\%$) achieved in both of these solutions is similar: peaking at around $\mathcal{D} = 30~\%$ for a composition of $c/c_{\mathrm{cmc}} = 1.7$ at a friction Reynolds number of $Re^* = 3000$. Therefore, if selecting or designing a viscoelastic surfactant solution for drag enhancement, tuning the extent of counter-ion is important; however, the total level of drag reduction that can be achieved at very high $Re$ is less sensitive to excess amounts of salicylate.

At a low TTAC concentration of $c/c_{\mathrm{cmc}} = 1.3$, slightly above the critical micelle concentration, it is clear from \figref{DR-1to5} that the fluid exhibits no drag enhancement and only drag reduction for friction Reynolds numbers of $Re^*>400$. Reducing the concentration to be $c/c_{\mathrm{cmc}} = 0.9$, below the critical micelle concentration, produces a drag curve that overlaps completely with measurement performed using water and no drag modification. This is similar to the results obtained using the $\xi=1$ TTAC solutions shown in \figref{DR-1to1} close to the critical micelle concentration. It is rare for drag reduction measurements at concentrations so close to the critical micelle concentration to be reported in the literature, but it appears that measuring the onset of drag reduction (or a lack thereof) is a robust way of identifying the critical micelle concentration -- at least for the present viscoelastic surfactant solutions with $\xi=1$ and 5.

\section{Discussion and Conclusions}

Surfactant solutions that form long wormlike micelles are a prototypical viscoelastic fluid that exhibit a variety of complex phenomena, many of which were demonstrated using our non-dilute solutions of the cationic surfactant trimethyltetradecylammonium chloride (TTAC) combined with sodium salicylate in deionized water. For example, non-dilute solutions of TTAC (with an equimolar ratio of sodium salicylate) were shown to exhibit (i) linear viscoelastic responses at low strains in good agreement with a single-mode Jeffreys model, (ii) the growth of normal stresses and onset of shear thinning at higher shear rates, (iii) large extensional relaxation times $\tau_E \sim \mathcal{O}(1~\mathrm{s})$ as measured using capillarity-driven extensional rheometry, and (iv) the onset of vorticity banding beyond a critical shear rate (measured using flow-induced birefringence in a parallel plate Rheo-Iris configuration). Such measurements provide a comprehensive rheological overview of the viscoelastic character of this cationic surfactant TTAC. 

\begin{figure*}[h!]
\centering
\includegraphics[width=\textwidth]{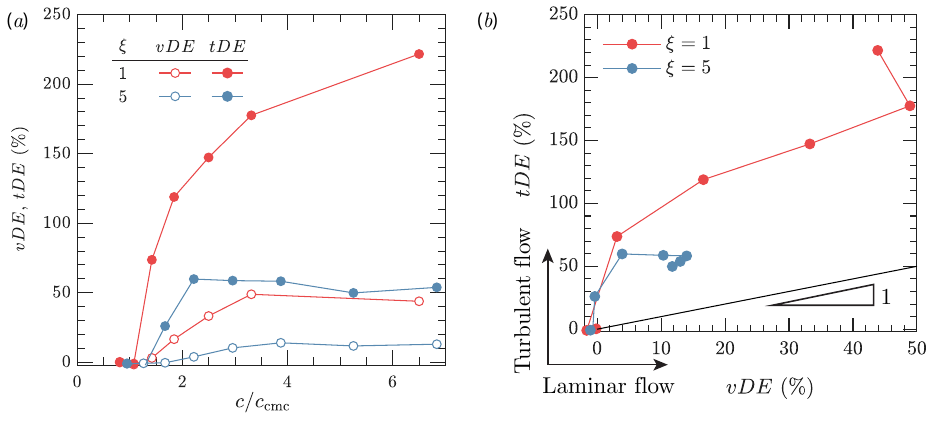}
\caption{(\textit{a}) Comparison of the viscous drag enhancement (open symbols) and turbulent drag enhancement (filled symbols) as a function of TTAC concentration for solutions with a relative salicylate counter-ion concentration of $\xi = 1$ and 5. For the TTAC solutions with $\xi=1$, viscous and turbulent drag enhancement are assessed for a viscous length scale of $\delta_{\nu}=\delta_{\mathrm{FIS}}=50~$\SI{}{\micro\meter} (equivalent to a friction Reynolds number of $Re^*_{\mathrm{FIS}} = 270$ in Taylor-Couette flow). For the $\xi = 5$ TTAC solutions, the drag enhancements are assessed for a viscous length scale of $\delta_{\mathrm{FIS}}=33~$\SI{}{\micro\meter} (equivalent to a friction Reynolds number of $Re^*_{\mathrm{FIS}} = 405$ in Taylor-Couette flow). (\textit{b}) Turbulent drag enhancement plotted with respect to viscous drag enhancement for the different flows of TTAC. The black solid line represents a parity line indicating when turbulent drag enhancement equals viscous drag enhancement, i.e., $tDE = vDE$.}
\label{DE}
\end{figure*}

When diluted into the dilute regime, the viscoelasticity of these TTAC solutions becomes difficult to measure directly but they were shown to develop flow-induced structures leading to the viscosity increasing abruptly at a critical shear rate. In a wide-gap Taylor-Couette flow, we have demonstrated that FIS formation triggers significant drag enhancement in the Taylor vortex flow regime. The extent of drag enhancement, and the critical conditions for onset, can be tuned based on the concentration of salicylate counter-ion. Larger counter-ion concentrations were found to delay the onset of drag enhancement to larger friction Reynolds numbers, and also diminish the extent of drag enhancement across all Reynolds numbers. We have also shown that drag enhancement under turbulent flow is triggered at conditions approximately equal to those required for the onset of flow-induced structures in viscometric flows. Kalliroscope flow visualization (cf \figref{TTCF}) revealed that the formation of flow-induced structures in Taylor-Couette flow produce spanwise-oriented G\"{o}rtler-like flow motions which disrupt the coherency of the turbulent Taylor vortices with increasing Reynolds numbers. 

Although it is apparent that the generation of flow-induced structures triggers drag enhancement in these sheared wormlike micelle solutions, the degree to which the shear-induced microstructure enhances viscosity in laminar shear flow (i.e., viscous drag modification) versus Taylor vortex flow (i.e., turbulent drag modification) differ significantly. In \figref{3iTT}(\textit{a}) we showed that flow-induced structures enhanced the viscosity of surfactant solutions by as much as $vDE = 50~\%$. However, in the Taylor-Couette flow, the same fluids produced a maximum drag enhancement of $tDE = 220~\%$, where the \textit{turbulent drag enhancement} is equivalent to the inverse of the drag modification defined in \eqnref{eqn:DM}, i.e., $tDE(Re^*) = -\mathcal{D}(Re^*)$. Clearly the flow kinematics and the differing response of the flow-induced microstructure in shear and extensional flow regimes plays a role in the dynamics of drag modification.

In \figref{DE}(\textit{a}) we present a direct comparison of the viscous drag enhancement (open symbols) and turbulent drag enhancement (filled symbols) for the TTAC solutions at different concentrations and two different counter-ion molar ratios of $\xi = 1$ and $\xi = 5$. Here the viscous and turbulent drag enhancement are compared in the viscometric laminar shear flow and in Taylor-Couette flow at a similar wall shear stress. Another way of interpreting this statement, is that the two flows are compared at comparable viscous length scales, given the viscous length scale depends on wall shear stress $\delta_\nu \propto \sigma_w^{-1/2}$. As we noted earlier, defining a viscous length scale is unconventional for a laminar shear flow, however this helps facilitate direct comparisons between laminar and turbulent flow conditions. For example, the drag enhancement attained with the TTAC solution (at a molar ratio of $\xi=1$) is evaluated for a viscous length scale of $\delta_{\mathrm{FIS}} = 50$~\SI{}{\micro\meter}. This corresponds roughly to the stress required to obtain maximum enhancement in the viscosity determined in the viscometric flow measurements of \figref{SIS}(\textit{b}), but also to the wall shear stress that generates the maximum turbulent drag enhancement in \figref{DR-1to1}(\textit{b}) -- given the corresponding friction Reynolds number is $Re^*_{\mathrm{FIS}} = d/\delta_{\mathrm{FIS}} = 270$. The same comparison is performed for the $\xi=5$ molar ratio TTAC solutions, but at a slightly smaller viscous length scale of $\delta_{\mathrm{FIS}} = 33$~\SI{}{\micro\meter} (i.e., a larger wall shear stress), and therefore, larger friction Reynolds number $Re^*_{\mathrm{FIS}} = 405$.

In \figref{DE}(\textit{b}) we show that for the same input stress, self-assembling surfactant solutions generate more shear resistance (drag) in turbulent Taylor-Couette flow than in laminar viscometric flow. The mixed kinematics in the Taylor-Couette flow (a combination of shear and extension) play a large role in generating additional contributions to the viscous drag. The degree of drag enhancement also depends significantly on the relative counter-ion concentration ($\xi$). Excessive amounts of salicylate produce lower levels of drag enhancement (viscous and turbulent); likely due to the lower probability of forming large unbranched micellar aggregates (which depend on persistence length of the wormlike micelles at equilibrium and the ionic strength of the solvent). Although images of the micelle microstructure for non-dilute TTAC solutions could be captured using cryo-TEM, images in the dilute regime are much more challenging to collect due to the limited contrast. One useful direction for future investigations would be better characterizing the microstructure of dilute wormlike surfactant solutions, and in particular, their microstructure before and after the onset of flow-induced structures and drag enhancement.

\begin{figure}[h!]
\centering
\includegraphics{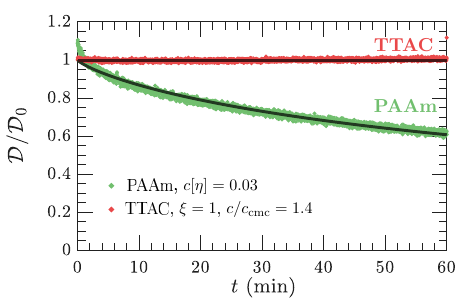}
\caption{Time-varying drag modification percentage of a dilute PAAm and TTAC solution. The PAAm concentration is $c[\eta] = 0.03$ (where the intrinsic viscosity is $[\eta] = 69.1~\mathrm{dl~g^{-1}}$), while the TTAC concentration is $c/c_{\mathrm{cmc}} = 1.4$ with a relative salicylate concentration of $\xi = 1$ (and a critical micelle concentration of $c_{\mathrm{cmc}} = 0.53~\mathrm{mM}$). For the TTAC solution, the drag reduction percentage remains relatively unchanged from its initial value of $\mathcal{D}_0 = 18.7~\%$. On the other hand, the PAAm solution exhibits a drag reduction percentage that decreases monotonically with respect to time, and a time-varying drag reduction percentage that is well described by the exponential function of \eqnref{eqn:mBrostow}, with an initial drag reduction percentage of $\mathcal{D}_0=12.6~\%$, breakage constant of $W = 3.7$, and a degradation timescale of $\tau_d = 356~\mathrm{min}$.}
\label{degrade}
\end{figure}

Although the onset and growth of drag enhancement at moderate shear rates is the most striking feature of these viscoelastic surfactant solutions, all test fluids (above the critical micelle concentration) also exhibit a large degree of drag reduction ($15~\%\lesssim \mathcal{D} \lesssim 30~\%$) at very high Reynolds numbers ($Re \gtrsim 50~000$). The amount of drag reduction depends on TTAC concentration, but is less sensitive to the counter-ion ratio (unlike the regime of drag enhancement). Importantly, this turbulent drag reduction is persistent over time, unlike the drag reduction achieved with synthetic polymers. To illustrate this, in \figref{degrade} we show the time-varying drag modification percentage in a strong Taylor-Couette flow at a bulk Reynolds number of $Re = 56~000$ for dilute solutions of PAAm and TTAC. Over the duration of 1~hour, the surfactant solution maintains a constant drag reduction percentage of $\mathcal{D}_0 = 18.7~\%$. On the other hand, irreversible scission of the PAAm molecules leads to a gradual decline in the extent of drag reduction over time. The time-varying drag reduction percentage of PAAm is accurately modeled using the exponential function proposed by \citet{Brostow1983},

\begin{equation}
    \frac{\mathcal{D}(t)}{\mathcal{D}_0} = \frac{1}{1+W\big(1-e^{-(t/\tau_d)} \big)}
    \label{eqn:mBrostow}
\end{equation}

\noindent
where $\mathcal{D}_0=12.6~\%$ is the initial drag reduction percentage of the PAAm solution at $t=0~\mathrm{min}$, $W = 3.7$ describes the number of breakage events per polymer chain, and $\tau_d = 356~\mathrm{min}$ is the characteristic timescale of degradation. After approximately $\Delta t = 60~\mathrm{min}$, the amount of drag reduction in the PAAm solution deteriorates to approximately $\mathcal{D} = 8.2~\%$, or 65~\% of its initial value.

In summary, we have demonstrated that the cationic surfactant TTAC is a robust turbulent drag modifier. Flow-induced alignment and concatenation of the wormlike micelles in these dilute TTAC solution yield appreciable drag enhancement, on top of the backdrop of time-varying Taylor vortex flow. However, at higher Reynolds numbers dilute solutions of the TTAC surfactant and salicylate counter-ion are also able to generate significant levels of drag reduction that are similar in magnitude to that achieved with dilute solutions of flexible synthetic polymers (such as PAAm), but without irreversible mechanical degradation. 

\section*{Supplemental Information}

The Supplemental Information contains three videos, in which we show the following results:

\begin{enumerate}
    \item A non-dilute wormlike micellar fluid undergoing capillarity-driven filament thinning, using the Dripping-onto-Substrate protocol. The aqueous wormlike micelle solution consists of TTAC at a concentration of $c = 17~\mathrm{mM}$ combined with NaSal at a molar ratio of $\xi=1.25$ (cf. \url{https://doi.org/10.6084/m9.figshare.34919400}).
    \item The time-evolving flow-induced birefringence of a wormlike micelle solution under viscometric shear flow (using a parallel plate geometry), collected using the Rheo-Iris configuration. The test fluid is an aqueous solution, comprised of an equimolar mixture of TTAC and NaSal ($\xi=1$), with a TTAC concentration of $c = 34$~mM. The Reynolds number of the flow is $Re_{\mathrm{pp}} = 0.16$ and the Weissenberg number is $Wi_R = 41.4$ (cf. \url{https://doi.org/10.6084/m9.figshare.34919451}).
    \item Kalliroscope visualization of a dilute wormlike micelle solution under strong shearing flow in the Taylor-Couette geometry. The solution is comprised of an equimolar mixture of TTAC and NaSal ($\xi=1$), with a TTAC concentration of $c = 0.74$~mM. The range of friction Reynolds numbers in the Taylor-Couette flow shown in the video is $50\leq Re^*\leq200$, which captures the onset of drag enhancement and appearance of G\"{o}rtler-like flow structures (cf. \url{https://doi.org/10.6084/m9.figshare.34919487}).
\end{enumerate}

\section*{Declarations}

The authors report no conflict of interest.

\section*{Acknowledgments}

LW and GHM acknowledge the support of the MIT Energy Initiative New Seed Innovation Fund. LW and GHM also recognize the support of Abhi Shetty and James Eikhoff with Anton Paar, Taisuke Sato and Taku Koizumi with Photonic Lattice, and Andrew Bridges with Photron, who were all instrumental in helping us commission, install and interpret the measurements collected using the Rheo-Iris/MCR-703 system discussed in \S~\ref{sec:semi-dilute}. The first author also acknowledges initial support to come to MIT from the Natural Sciences and Engineering Research Council of Canada (NSERC), [PDF-587339-2024]. Cette recherche a \'{e}t\'{e} financ\'{e}e par le Conseil de recherches en sciences naturelles et en g\'{e}nie du Canada (CRSNG), [PDF-587339-2024]. 

\section*{Data Availability}
 
The data that support the findings of this study are available from the corresponding author upon reasonable request.

\appendix

\section{Polyacrylamide}\label{app:PAAm}

For comparative purposes, aqueous solutions of a synthetic drag-reducing polymer are also prepared for measurements in the Taylor-Couette geometry, depicted in \figref{TTCF}(\textit{a}). The chosen synthetic polymer was an anionic polyacrylamide (PAAm) manufactured by SNF Floerger (with a product identifier of 6030S). An initial solution with a concentration of $c = 0.1~\mathrm{g~dl}^{-1}$ was prepared by dissolving 1~g of solid polymer powder into 10~dl of deionized water, along with $100~\mathrm{mM}$ of the divalent salt, magnesium sulfate (with a molar mass of $120.37~\mathrm{g~mol^{-1}}$). Divalent ions were added to screen charge interactions along the backbone of the polyelectrolyte PAAm chain, and mitigate intra- and intermolecular associations \citep{Shetty2009}. Seven solutions of different PAAm concentration were then prepared by successively diluting the concentrated master solution with solvent (i.e., $100~\mathrm{mM}$ of magnesium sulfate in deionized water). The seven PAAm concentrations included: $c = 0.0005, 0.001, 0.002, 0.005, 0.01, 0.02$ and 0.05~g~dl$^{-1}$. 

\begin{figure*}[h!]
\centering
\includegraphics[width=\textwidth]{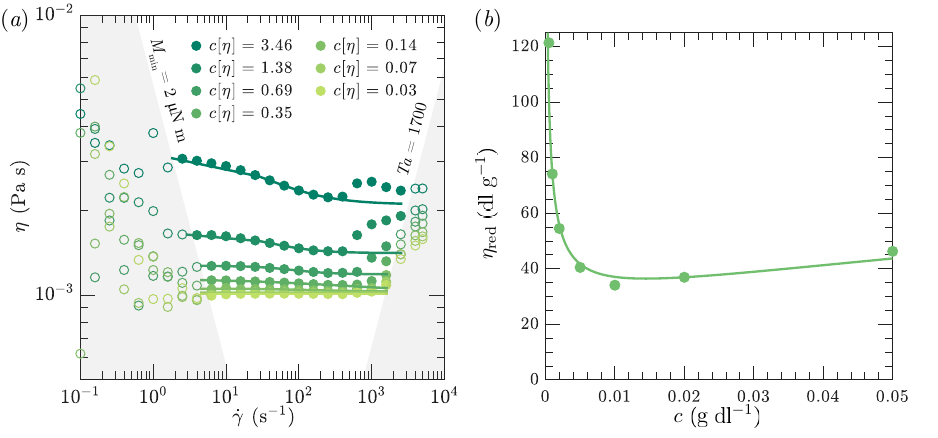}
\caption{Steady shear viscosity measurements shown in (\textit{a}) of seven solutions of polyacrylamide in a solvent comprised of deionized water and 100~mM of magnesium sulfate. Solid coloured lines represent fits to the shear rheograms using the Cross model of \eqnref{eqn:Cross}. Measurements of the reduced viscosity $\eta_{\mathrm{red}}=(\eta_0-\eta_s)/\eta_sc$ as a function of PAAm concentration are shown in (\textit{b}). The green solid line in (\textit{b}) represents fits using the empirical model of \eqnref{eqn:Fuoss}, with a Fuoss coefficient of $\mathcal{F} = 2.2~\mathrm{dl^{1/2}~g^{-1/2}}$ and an intrinsic viscosity of $[\eta] = 69.1~\mathrm{dl~g^{-1}}$.}
\label{PAAm-visc}
\end{figure*}

Steady shear viscosity measurements, $\eta(\dot{\gamma})$, of the PAAm solutions at different concentrations are shown in \figref{PAAm-visc}(\textit{a}). Shear viscosity measurements were collected using the same torsional rheometer (DHR-3, TA Instruments) and double gap geometry as before (cf. \figref{SIS}). At high PAAm concentrations, the solutions exhibit shear-thinning rheology and a zero-shear--rate viscosity that is approximately three to four times larger than the solvent ($\eta_s = 0.95~\mathrm{mPa~s}$). The largest PAAm concentrations exhibit an apparent shear thickening, at rates below the onset of Taylor vortex flow $\dot{\gamma}\approx 400~\mathrm{s}^{-1}$ -- with a similar appearance to the flow-induced structures seen in \figref{SIS}(\textit{a}). However, we believe this arises from elastic instabilities (driven by elastic hoop stresses), given these solutions also exhibit significant levels of elasticity (as measured using capillary thinning extensional rheometry). On the other hand, the dilute surfactant solutions discussed in \S~\ref{sec:SIS} show no measurable elasticity from capillary thinning experiments, something also observed in the investigation by \citet{Warwaruk2021,Warwaruk2023}. As the solution is diluted, the viscosity decreases in magnitude and exhibits less shear-thinning. The shear-thinning rheology of each test fluid is fit using the model of \citet{Cross1965}

\begin{equation}
    \frac{\eta(\dot{\gamma})-\eta_\infty}{\eta_0-\eta_\infty}=\frac{1}{1+(\dot{\gamma}/\dot{\gamma}_0)^m}
    \label{eqn:Cross}
\end{equation}

\noindent
where $\eta_0$ is the zero-shear--rate viscosity, $\eta_{\infty}$ is the infinite-shear--rate viscosity, $\dot{\gamma}_0$ is the shear rate at which the viscosity transitions from a constant $\eta_0$ to shear thinning, and $m$ is the power law index. For brevity, the fitted results are not listed here; however, to determine the intrinsic viscosity of the PAAm polymer, fitted values of the zero-shear--rate viscosity are used to determine the reduced viscosity of each solution according to $\eta_{\mathrm{red}} = (\eta_0-\eta_s)/\eta_sc$.

\begin{figure*}[h!]
\centering
\includegraphics[width=\textwidth]{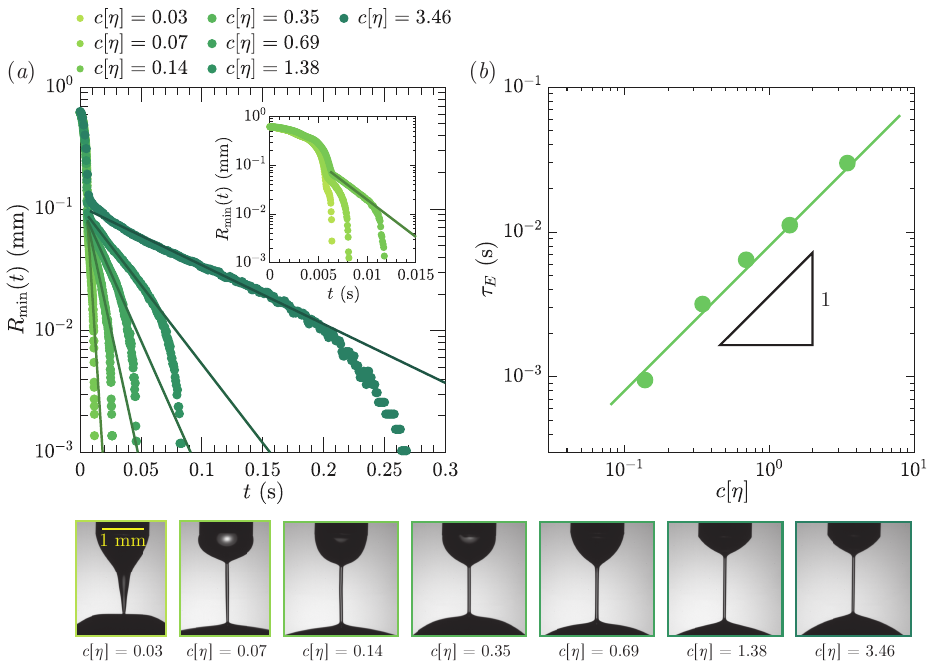}
\caption{Measurements of (\textit{a}) the time-varying minimum filament radius $R_{\mathrm{min}}(t)$ obtained from DoS extensional rheometry of the seven PAAm solutions (with an intrinsic viscosity of $[\eta] = 69.1~\mathrm{dl~g^{-1}}$). Coloured lines in (\textit{a}) represent fits of the elasto-capillary regime using the exponential expression given in \eqnref{eqn:EC}. The inset axis in (\textit{a}) shows the time-varying filament radius for the three lowest PAAm concentrations. Extensional relaxation times (estimated from fits to the elastocapillary regime) as a function of PAAm concentration are shown in (\textit{b}). The black solid line in (\textit{b}) depicts a linear scaling of the relaxation time with concentration, i.e., $\tau_E \propto c$. Instantaneous snapshots of the filament radius are shown in the bottom panels of the figure. Images are taken at different time instances, but a similar filament radius $R_{\mathrm{min}} = 0.04~\mathrm{mm}$}
\label{PAAm-relax}
\end{figure*}

Measurements of the reduced viscosity as a function of PAAm concentration are shown in \figref{PAAm-visc}(\textit{b}). The reduced viscosity exhibits a nonmonotonic trend, first decreasing with concentration for $c\leq 0.01~\mathrm{g~dl^{-1}}$ then increasing for concentrations of $c > 0.01~\mathrm{g~dl^{-1}}$. The increase in reduced viscosity with dilution is typical of polyelectrolytes and is attributed to changes in polymer conformation caused by electrostatic intramolecular repulsions \citep{Fuoss1948}. This is despite the addition of divalent ions to the solution. A modified version of the empirical equation proposed by \citet{Fuoss1948} is fit to measurements of the reduced viscosity as a function of concentration

\begin{equation}
    \eta_{\mathrm{red}}(c) = \frac{\mathcal{F}}{\sqrt{c}}\Big(1+[\eta]c +\mathcal{O}(c^2) \Big)
    \label{eqn:Fuoss}
\end{equation}

\noindent
where $\mathcal{F}$ is denoted the \textit{Fuoss} coefficient, describing the degree of polyelectrolyte interactions at low concentrations, and $[\eta]$ is an estimate for the intrinsic viscosity. Here, the intrinsic viscosity corresponds to the local minimum in the reduced viscosity predicted from \eqnref{eqn:Fuoss}, i.e., $c_{\mathrm{min}}=[\eta]^{-1}$. At low concentrations $c<[\eta]^{-1}$, the reduced viscosity decreases with increasing PAAm concentration according to the scaling proposed by \citet{Fuoss1948}, i.e., $\eta_{\mathrm{red}}\propto c^{-1/2}$. At high concentrations $c\geq [\eta]^{-1}$, inter-chain interactions cause the reduced viscosity to increase and deviate from the Fuoss scaling (as described using a Taylor series expansion). The empirical model of \eqnref{eqn:Fuoss}, ignoring the higher-order terms $\mathcal{O}(c^2)$, accurately describes the reduced viscosity for the PAAm solutions with increasing concentration, and is represented by the green solid line in \figref{PAAm-visc}(\textit{b}). The resulting Fuoss coefficient is $\mathcal{F}~=~2.2~\mathrm{dl^{1/2}~g^{-1/2}}$ and the intrinsic viscosity is $[\eta] = 69.1~\mathrm{dl~g^{-1}}$. This estimate of the intrinsic viscosity is qualitatively supported by the results presented in \figref{PAAm-visc}(\textit{a}), which show that the semi-dilute solutions of PAAm (with $c[\eta]>1$) are shear thinning and exhibit larger overall viscosities compared to the dilute solutions.

\begin{figure*}[h!]
\centering
\includegraphics[width=\textwidth]{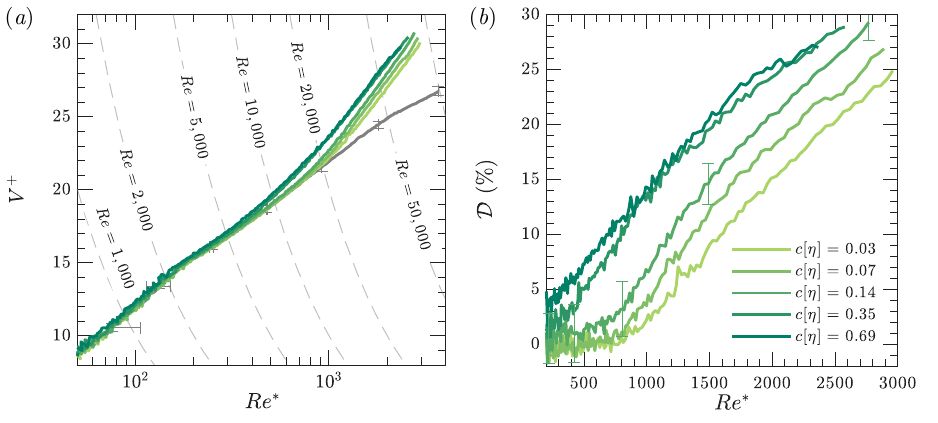}
\caption{Measurements of the skin-friction drag curves, presented in Prandtl-von K\'arm\'an coordinates, $V^+(Re^*)$, are shown in (\textit{a}) for solutions with the five lowest PAAm concentrations. The grey solid line in (\textit{a}) represent measurements with a Newtonian fluid (water). Grey dashed lines in (\textit{a}) are nomograms of constant bulk Reynolds number $Re\equiv Re^*V^+$. The resulting drag reduction percentage, established using \eqnref{eqn:DM}, as a function of friction Reynolds number $Re^*$ is shown in (\textit{b}) for the PAAm solutions at different concentrations.}
\label{PAAm-DR}
\end{figure*}

Dripping-onto-substrate extensional rheometry is used to measure the extensional relaxation time of the PAAm solutions at different concentrations. The same experimental configuration discussed in \S~\ref{sec:ion} is employed for the current measurements. \figref{PAAm-relax}(\textit{a}) shows measurements of the time-varying filament radius for the PAAm solutions. All solutions with concentrations of $c[\eta]\geq0.35$ exhibit a well-defined elasto-capillary regime and minimum filament radius that decays exponentially with time according to \eqnref{eqn:EC}. Sample images of the filament radius, shown in the bottom panel of \figref{PAAm-relax}, also confirm that the filament is slender and cylindrical within the regime of exponential thinning. Fits of \eqnref{eqn:EC} to the measurements of the time-varying filament radius are represented by the solid coloured lines in \figref{PAAm-relax}(\textit{a}), and used to extract an estimate of the extensional relaxation time, $\tau_E$. Fits using \eqnref{eqn:EC} to the two solutions with the lowest PAAm concentrations were not performed, considering the solutions did not exhibit a long-lived exponential thinning regime, and imaging showed a markedly non-cylindrical filament (as shown in the instantaneous images at the lower left of \figref{PAAm-relax}).  Measured values of the relaxation time are plotted with respect to PAAm concentration in \figref{PAAm-relax}(\textit{b}). The relaxation time exhibits an approximately linear variation with concentration. \citet{Soetrisno2023} similarly observed a linear scaling between the relaxation time of solutions comprised of high molecular weight PAAm polymers with a large degree of polydispersity (similar to the present PAAm polymers).

The five dilute solutions of PAAm with concentrations of $0.03\leq c[\eta]\leq0.69$, were also tested in the wide-gap Taylor-Couette geometry depicted in \figref{TTCF}(\textit{a}). The same procedure detailed in \S~\ref{sec:TTCF} was used to measure the drag curves of the PAAm solutions. The angular velocity of the rotor was gradually increased from $\varOmega = 1~\mathrm{rad~s^{-1}}$ to $\varOmega = 270~\mathrm{rad~s^{-1}}$ over a duration of $11~\mathrm{min}$, at a slow (logarithmic) growth rate of $9\times10^{-3}~\mathrm{s}^{-1}$. The resulting drag curves, plotted in Prandtl-von K\'arm\'an coordinates of $V^+(Re^*)$ are shown in \figref{PAAm-DR}(\textit{a}). In contrast to the surfactant solutions, cf. \figref{DR-1to1} and \figref{DR-1to5}, none of the PAAm solutions exhibit drag enhancement at low to moderate Reynolds numbers. At a sufficiently large Reynolds number, in the range of $300\leq Re^*\leq 900$ depending on PAAm concentration, the PAAm flows enter a state of drag reduction, in which values of dimensionless rotor velocity begin to exceed the Newtonian flow of water. In \figref{PAAm-DR}(\textit{b}) we show the resulting values of the drag modification percentage as a function of the friction Reynolds number $\mathcal{D}(Re^*)$ -- here focusing on positive values that correspond to drag reduction. As the PAAm concentration increases, so does the level of drag reduction for a given Reynolds number. However, the extent of drag reduction appears to saturate or asymptote for the highest concentrations.

To quantify the degree to which polymers are extended within the Taylor-Couette flow, a bulk Weissenberg number is defined as $Wi = \tau_E \varOmega R_i/d$, where $\varOmega R_i/d$ is the average shear rate across the gap of the Taylor-Couette flow. Taking a ratio of the Weissenberg to Reynolds number yields the elasticity number $El = Wi/Re = \tau_E \eta_0/ \rho d^2$, a dimensionless quantity that is independent of the rotor speed and characterizes the relative importance of elastic and inertial forces in the viscoelastic flows of PAAm. Using the measured values of $\eta_0$ from \figref{PAAm-visc} and $\tau_E$ from \figref{PAAm-relax}, the elasticity number for the three most concentrated PAAm solutions shown in \figref{PAAm-DR} with concentrations of $c[\eta] = 0.14$, $0.35$ and $0.69$ are found to be $El = 6.3\times10^{-6}$, $1.6\times10^{-5}$, and $3.4\times10^{-5}$, respectively. These values of the elasticity number are not significantly different than those observed for dilute solutions of natural salivary mucins with $1.2\times10^{-5}\leq El\leq5.0\times10^{-5}$ determined from our previous work \citep{Warwaruk2026b}, which achieved similar degrees of drag reduction ($\mathcal{D}\approx 30~\%$ for $Re^* = 2000$).

\bibliographystyle{bib_style} 
{\footnotesize \bibliography{cas-refs}}

\end{document}